# A practical guide to machine learning interatomic potentials – Status and future

**Authors:** Ryan Jacobs[1,*], Dane Morgan[1,*], Siamak Attarian[1], Jun Meng[1], Chen Shen[1], Zhenghao Wu[2], Clare Yijia Xie[3], Julia H. Yang[2,3], Nongnuch Artrith[4], Ben Blaiszik[5,6], Gerbrand Ceder[7,8], Kamal Choudhary[9], Gabor Csanyi[10], Ekin Dogus Cubuk[11], Bowen Deng[7,8], Ralf Drautz[12], Xiang Fu[13], Jonathan Godwin[14], Vasant Honavar[15,16,17,18], Olexandr Isayev[19,20], Anders Johansson[3], Boris Kozinsky[3], Stefano Martiniani[21,22,23], Shyue Ping Ong[24], Igor Poltavsky[25], KJ Schmidt[5,6], So Takamoto[26], Aidan Thompson[27], Julia Westermayr[28], Brandon M. Wood[13]

**Affiliations:**

[1]Department of Materials Science and Engineering, University of Wisconsin-Madison, Madison, WI, USA, 55705.

[2]Harvard University Center for the Environment, Harvard University, Cambridge, MA 02138, USA

[3]John A. Paulson School of Engineering and Applied Sciences, Harvard University, Cambridge, MA 02138, USA

[4]Materials Chemistry and Catalysis, Debye Institute for Nanomaterials Science, Utrecht University, Utrecht 3584 CG, The Netherlands

[5]Globus, University of Chicago, Chicago, IL, United States of America

[6]Data Science and Learning Division, Argonne National Laboratory, Lemont, IL, United States of America

[7]Department of Materials Science and Engineering, University of California, Berkeley, California 94720, United States

[8]Materials Sciences Division, Lawrence Berkeley National Laboratory, California 94720, United States

[9]Material Measurement Laboratory, National Institute of Standards and Technology, Gaithersburg, MD 20899, USA

[10]Department of Engineering, University of Cambridge, Cambridge, CB2 1PZ, UK

[11]Google DeepMind, Mountain View, CA, USA

[12]Interdisciplinary Centre for Advanced Materials Simulation (ICAMS), Ruhr-Universität Bochum, 44780 Bochum, Germany

[13]Fundamental AI Research (FAIR) at Meta, USA

[14]Orbital Materials, London, England, United Kingdom

[15]Department of Computer Science and Engineering, The Pennsylvania State University, University Park, PA, USA.

[16]College of Information Sciences and Technology, The Pennsylvania State University, University Park, PA, USA

[17]Artificial Intelligence Research Laboratory, The Pennsylvania State University, University Park, PA, USA

[18]Center for Artificial Intelligence Foundations and Scientific Applications, The Pennsylvania State University, University Park, PA, USA

[19]Department of Chemistry, Mellon College of Science, Carnegie Mellon University, Pittsburgh, Pennsylvania 15213, United States

[20]Computational Biology Department, School of Computer Science, Carnegie Mellon University, Pittsburgh, Pennsylvania 15213, United States

[21]Courant Institute of Mathematical Sciences, New York University, New York, NY 10003

[22]Center for Soft Matter Research, Department of Physics, New York University, New York, NY 10003

[23]Simons Center for Computational Physical Chemistry, Department of Chemistry, New York University, New York, NY 10003

[24]Aiiso Yufeng Li Family Department of Chemical and Nano Engineering, University of California, San Diego, 9500 Gilman Dr. Mail Code #0448, La Jolla, CA 92093

[25]Department of Physics and Materials Science, University of Luxembourg, L-1511 Luxembourg, Luxembourg

[26]Preferred Networks, Inc., Otemachi Bldg., 1-6-1 Otemachi, Chiyoda-ku, Tokyo, 100-0004, Japan

[27]Center for Computing Research, Sandia National Laboratories, New Mexico, United States of America

[28]Wilhelm-Ostwald-Institut für Physikalische und Theoretische Chemie, Universität Leipzig, Germany

**\*Corresponding Author e-mails:** rjacobs3@wisc.edu, ddmorgan@wisc.edu

**Abstract**

The rapid development and large body of literature on machine learning interatomic potentials (MLIPs) can make it difficult to know how to proceed for researchers who are not experts but wish to use these tools. The spirit of this review is to help such researchers by serving as a practical, accessible guide to the state-of-the-art in MLIPs. This review paper covers a broad range of topics related to MLIPs, including (i) central aspects of how and why MLIPs are enablers of many exciting advancements in molecular modeling, (ii) the main underpinnings of different types of MLIPs, including their basic structure and formalism, (iii) the potentially transformative impact of universal MLIPs for both organic and inorganic systems, including an overview of the most recent advances, capabilities, downsides, and potential applications of this nascent class of MLIPs, (iv) a practical guide for estimating and understanding the execution speed of MLIPs, including guidance for users based on hardware availability, type of MLIP used, and prospective simulation size and time, (v) a manual for what MLIP a user should choose for a given application by considering hardware resources, speed requirements, energy and force accuracy requirements, as well as guidance for choosing pre-trained potentials or fitting a new potential from scratch, (vi) discussion around MLIP infrastructure, including sources of training data, pre-trained potentials, and hardware resources for training, (vii) summary of some key limitations of present MLIPs and current approaches to mitigate such limitations, including methods of including long-range interactions, handling magnetic systems, and treatment of excited states, and finally (viii) we finish with some more speculative thoughts on what the future holds for the development and application of MLIPs over the next 3-10+ years.

## 1 Introduction

This paper was inspired by the workshop "Machine Learning Potentials – Status and Future (MLIP-SAFE)", which was held online on July 17-19, 2023. It represents select themes and key points we thought would be of particular interest to the broader materials science and chemistry communities. The rapid development and large body of literature on machine learning potentials (MLIPs, sometimes also called machine learning force fields or MLFFs) can make it difficult to know how to proceed for researchers who are not experts but wish to use these tools. The spirit of this paper is to help such researchers by serving as a practical, accessible guide to the state-of-the-art in MLIPs. We aim to keep deep mathematics and formalism to a minimum, as such details can be readily found in other excellent reviews and references therein.[1–9] In contrast, we believe that guidance on the general landscape of MLIPs, their practical use, trade-offs, pros and cons for particular problems, timing, and how to get started running them is still challenging to learn from the literature. We note that a recent Comment published by Ko and Ong and a Perspective from Duignan both highlight many of the same topics addressed in this Review,[10,11] albeit more briefly and at a higher level, and we therefore feel the present work serves as a complementary, more in-depth examination of the present state of MLIPs from a practical perspective. Our target audience is technically literate material scientists and chemists, with a background in molecular modeling, but not MLIP experts. Therefore, we do not discuss technical details of, for example, basis function expansions, but do provide guidance on how to understand the broad differences between approaches (e.g., atomic cluster expansion (ACE) vs. graph neural networks (GNNs)) and the benefits and tradeoffs of using different approaches. This paper will provide a high-level guide on the key fundamental aspects needed to understand the landscape of MLIPs, including their enormous potential range of applications, general frameworks, typical workflows (including fitting and/or using pre-fit MLIPs), speed and accuracy, supporting infrastructure, and some guidance on MLIP choice.

This review paper covers a broad range of topics related to MLIPs and is organized as follows. In **Sec. 2**, we provide a list of MLIPs discussed throughout this review, including their abbreviations and key references to original work. In **Sec. 3**, we outline the central aspects of how and why MLIPs are enablers of many exciting advancements in molecular modeling. In **Sec. 4**, we discuss the main underpinnings of different types of MLIPs, including their basic structure and formalism (**Sec. 4.1**), the differences between MLIPs using explicit featurization approaches of the atomic environments vs. implicit approaches leveraging graph neural networks (**Sec. 4.2**) and details of the explicit and implicit approaches more specifically in **Sec. 4.3** and **Sec 4.4**, respectively. In **Sec. 5**, we highlight the potentially transformative impact of universal MLIPs (U-MLIPs) for both organic and inorganic systems, including an overview of the most recent advances, capabilities, downsides, and potential applications of this nascent class of MLIPs. In **Sec. 6**, we provide a practical guide for estimating and understanding the execution speed of MLIPs, including guidance for users based on hardware availability, type of MLIP used, and prospective simulation size and time. Next, **Sec. 7** functions as a practical manual for what MLIP a user should choose for a given application by considering hardware resources (**Sec. 7.1**), speed requirements (**Sec. 7.2**), energy and force accuracy requirements (**Sec. 7.3**), as well as guidance for choosing pre-trained potentials (**Sec. 7.4**), and fitting a new potential from scratch (**Sec. 7.5** and **Sec. 7.6**). Discussion in **Sec. 8** centers around MLIP infrastructure, including sources of training data, pre-trained potentials, and hardware resources for training. **Sec. 9** summarizes some key limitations of present MLIPs and current approaches to mitigate such limitations, including methods of including long-range interactions, handling magnetic systems, and treatment of excited states. Finally, we conclude in **Sec. 10** with some more speculative thoughts on what the future holds for the development and application of MLIPs over the next 3-10+ years.

## 2   A List of MLIPs

In the following discussions, we will often refer to MLIPs by their acronyms. To help clarify the meaning and appropriate citations for these MLIPs we here summarize the names, acronyms, and standard citations of the MLIPs that are discussed in this paper. Note that this is not meant to serve as a comprehensive list of existing MLIPs.

Accurate NeurAl networK engINe for Molecular Energies (ANAKIN-ME, ANI for short): [12]
Allegro: [13]
Atomic Cluster Expansion (ACE): [14]
Atomic Energy Network (ænet): [15,16]
Atomistic Line Graph Neural Network-based Force Field (ALIGNN-FF): [17]
Atoms-In-Molecules Network 2 (AIMNet2): [18]
Behler-Parrinello Neural Network (BP-NN, or BP)[19]
Crystal Hamiltonian Graph Neural Network (CHGNet)[20]
Deep Molecular Dynamics (DeepMD): [21,22]
Elemental Spatial Density Neural Network Force Field (Elemental-SDNNFF): [23]
EquiformerV2-OMAT24: [24]
Fast Learning of Atomistic Rare Events (FLARE): [25]
Gaussian Approximation Potential (GAP): [26]
Graph-based Pre-trained Transformer Force Field (GPTFF): [27]
Graph Networks for Materials Exploration (GNoME): [28]
Graph Atomic Cluster Expansion (grACE):  [29]

Mattersim: [30]

ACE with message passing (MACE): [31]

MACE foundation model (MACE-MP-0): [32]

MACE-OFF23 potential for organics (MACE-OFF23): [33]

Moment Tensor Potential (MTP): [34]

Neural Equivariant Interatomic Potential (NequIP): [35]

Orb: [36]

PreFerred Potential (PFP): [37]

Scalable EquiVariance-Enabled Neural NETwork (SevenNet): [38]

SchNet: [39]

Spectral Neighbor Analysis Potential (SNAP): [40]

Three-body Materials Graph Network (M3GNet): [41]

Ultra-Fast Force Fields (UF3) potential: [42]

## 3   What Makes MLIPs So Exciting?

For this paper, we will define an MLIP as a function that takes as input a set of atoms with positions $\{x_i, y_i, z_i\}$ (optionally including a set of periodic lattice vectors), and element types $\{n_i\}$ and maps this atomic configuration to a total energy $E$ for that set of atoms $i$. The MLIP therefore serves as a potential energy surface (PES) function. The MLIP generally also provides forces (and stresses in cases with periodic boundary conditions), which are spatial derivatives of the PES generated by the MLIP. The forces and stresses are generally available through an analytical derivative expression that can be obtained from the MLIP and no numerical differentiation of $E$ is required. (We note that some of the presently best-performing MLIPs are trained separately on energies and forces, and are non-conservative in the sense that the forces are not directly calculated by differentiating the PES[24,30,36].) The purpose of an MLIP is to enable efficient calculation of material properties, typically using molecular dynamics (MD), for myriad applications ranging from understanding and predicting chemical reactions to designing stronger metal alloys to developing more effective drugs. We note that here we define a "material" to mean any collection of atoms, from crystals to gasses to molecules. Throughout this work, we consider a model to be an MLIP if it can provide energies and its gradients.

Historically, atomistic simulation of materials has been divided into two very different approaches. On the one hand, *ab initio* molecular dynamics (AIMD) has enabled high accuracy simulations of small numbers of atoms, providing rich insight into the structural, thermodynamic, and transport properties of materials at the very smallest scales. On the other hand, molecular dynamics simulations with physics-based potentials (PBPs) have enabled researchers to qualitatively study how atomic interactions drive the emergence of diverse phenomena on much larger scales. For a long time, these two approaches were disconnected. AIMD was incapable of achieving the scale needed to observe many phenomena of scientific interest, while PBP-based MD could not provide accurate enough representations of any specific material. The emergence of MLIPs has revolutionized the practice of atomistic simulations by bridging this disconnect. By leveraging massively parallel computing resources and flexible parallel simulations frameworks such as LAMMPS,[43] it is now possible to directly simulate large-scale emergent phenomena in specific materials with accuracy that approaches that of AIMD.

MLIPs differ from traditional PBPs in that MLIPs utilize a highly flexible approach to represent the PES function (e.g., a neural network), typically taken from the machine learning (ML) community. In contrast, PBPs use a highly constrained functional form guided by physical intuition (e.g., a Lennard-Jones or Born-Meyer potential). The categories of PBPs vs. MLIPs are somewhat arbitrary and inexact, as there is really a continuum of possible approaches between the extreme limits of a functional forms derived from physics with almost no fitting parameters (a pure PBP) and a purely numerical fit done with almost no intuitive guidance (a pure MLIP). An overview of the different general approaches for constructing PBPs and MLIPs is provided in **Figure 1**. Starting from the simple limit, PBPs can incorporate increasingly flexible functions to become more like ML models, e.g., as has been done in the very flexible forms for pair interactions in the Embedded Atom Method (EAM) potentials.[44,45] Conversely, starting from the pure ML side, MLIPs can be made more like PBPs by introducing physically-motivated terms to the PES representation, e.g., adding in a Ziegler-Biersack-Littmark repulsive interaction to ensure that atoms do not behave unphysically when close together, as is available in several MLIP training packages.[21,22,46] In addition, many intermediate approaches are possible, e.g., as discussed in the review by Mishin.[4] Here, we will follow the standard convention of referring to any potential that uses traditional ML featurization or modeling approaches as an MLIP.

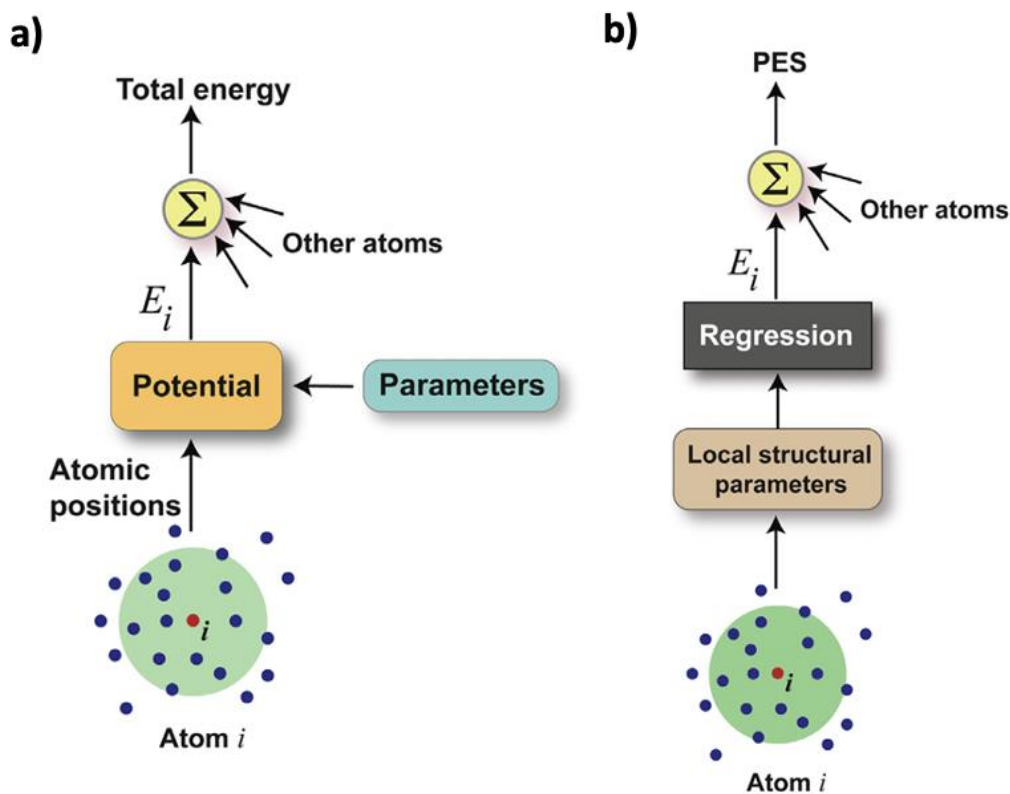

**Figure 1.** Overview of approaches for generating (a) physics-based potentials and (b) machine learning-based potentials. Adapted with permission from Ref. [47].

MLIPs have an advantage vs. PBPs because their flexible functional form can fit essentially arbitrarily complex atomistic scale potential energy landscapes. We note that by "energy landscape" we mean the ground state Born-Oppenheimer surface, as is generally produced by *ab initio* calculations. One of the main disadvantages of MLIPs vs. PBPs is that MLIPs require a lot of training data to learn the physics of the system. However, as *ab initio* data continues to become more plentiful, more accurate, easier to obtain, and standardized in publicly available databases (e.g., the MPtrj database[20] contained in the Materials Project and the Open Materials 24 (OMAT24) database released by Meta[24]) the high training data requirements of MLIPs become increasingly easy to meet, giving MLIPs a notable and increasing advantage over PBPs. We can think of MLIPs today as an improved version of traditional PBPs, but with greater accuracy and more flexibility to model complex systems, at the expense of higher computational cost (depending on the type of MLIP used). Very flexible and accurate PBP functional forms are often difficult to develop because they require significant domain expertise and physical insight to construct.

When using MLIPs, it is important to state that the level of improvement in accuracy and number of different elements modeled appears to be so great in comparison with PBPs that the introduction of MLIPs appears more revolutionary than evolutionary. The functional forms in PBPs, for all their ingenuity, almost always do not have sufficient complexity to quantitatively model the behavior of interacting atoms across all the conditions of interest, which often involve many complex changes in bonding and charge state. In contrast, modern MLIPs can capture many chemical changes provided adequate training data is available. We stress that MLIPs are not fundamentally limited in any particular way, e.g., to only metallic or ionic systems, or to only nonreactive vs. chemically reactive processes. While this is a good initial perspective for those new to MLIPs there are definitely some constraints on present MLIP capabilities, and we enumerate some of the major present limitations of MLIPs in **Sec. 9**. Distinctions that were often essential to determining the form and applicability of PBPs, e.g., organics vs. inorganics, bond-breaking / reactive vs. not, metallic vs. ionic vs. covalent, are often not particularly important for whether an MLIP is applicable. Furthermore, the accuracy of MLIPs is typically on the scale of a few to tens of meV/atom, which is often an order of magnitude better than typical PBPs.[1,48] Additionally, MLIPs are straightforward to iteratively improve and can be fixed if they show undesirable errors by adding more training data.[49] While sometimes PBPs can be iteratively improved as well, doing so is more difficult than improving MLIPs, because instead of just providing more diverse training data, more fundamental changes to the underlying functional forms may be needed, which requires significant expertise to do properly. Finally, MLIPs with excellent testing errors are quite easy to fit (typically ranging from just days to a couple of months for a graduate student with the necessary skills for a system comprising a few elements), and good pre-trained potentials, including ones covering large parts of chemical and structural space, are becoming widely available, e.g., as seen with the recent development of Universal MLIPs (U-MLIPs) (see **Sec. 5**). Given all the advantages of MLIPs, it seems possible that MLIPs will be easy enough to train for most systems that they may at least partially replace *ab initio* calculations in applications needing just forces and energies. Even partial replacement of *ab initio* calculations will dramatically accelerate many kinds of molecular modeling, but one notable example is that quantum mechanics-based AIMD might be almost entirely replaced by MLIP MD. This replacement of AIMD with MLIP MD (and analogously for geometry optimization) would make similar time and length scales to those studied with AIMD accessible with orders of magnitude less compute time. Such an increased efficiency is an important change as a significant amount of the compute time used in *ab initio* simulations is devoted to running AIMD (e.g., in a recent year approximately 40% of the UK's academic HPC was spent on DFT calculations, most of which was AIMD). Perhaps more importantly, the use of MLIP MD would unlock gains of orders of magnitude in accessible length and time scales vs. AIMD, enabling the study of new physical regimes inaccessible with AIMD. There is growing evidence that we will be able to develop quantitative U-MLIPs, something like the foundational models in computer vision and language machine learning, which can directly, or with some fine tuning, provide almost instant access to quantum mechanical accuracy on almost any chemical system at the scale of millions to billions of atoms and for microsecond or longer timescales.[20,28,32,41] Thus, MLIPs may dramatically enhance the capabilities of molecular simulations, significantly impacting chemistry, biology, materials science and engineering, physics, and many other disciplines. The necessary understanding, methods, and tools exist today to enable non-experts to apply MLIPs to practical problems, and it is reasonable to expect an explosion of use across many fields of science in the next few years. However, there are still significant challenges to realizing the full potential of MLIPs, including refining the best features and architectures, developing optimal training strategies, finding ways to include additional physics (e.g., long-range interactions), scaling up to universal potentials, and successfully developing and adopting potentials for many complex systems of interest.

## 4  Understanding the Types of MLIPs - Basic Formalisms

In this section, we discuss the basic formalism behind MLIPs. The goal of this discussion is to provide a qualitative description to help guide users in understanding what aspects control the key properties users care about, which include e.g., (1) human vs. computational MLIP training limitations, (2) MLIP speed of execution, (3) MLIP accuracy, (4) MLIP ease of use, and (5) appropriateness of an MLIP to specific problems. Detailed mathematical descriptions of MLIP formalisms can be found in many other reviews.[1–9] This section provides a high-level overview of the basic construction of an MLIP (**Sec. 4.1**), discussion of the construction and use cases of MLIPs created by explicitly featurizing atomic positions with specific functional forms (**Sec. 4.3**), discussion of the construction of MLIPs created implicitly through featurizing by graph neural network approaches (**Sec. 4.4**), the general differences between these two approaches (**Sec. 4.2**), and, finally, the unification of these two approaches into a single over-arching MLIP framework (**Sec. 4.5**). We stress that this section was written to reflect the historical development of different MLIP formalisms, where we discuss differences between various approaches which we believe accurately portrays how the community has thought of MLIP development until recently. However, these previously perceived differences between various MLIP formalisms appears to be collapsing into a single over-arching formalism, which we discuss in more detail in the following subsections.

## 4.1 The Basic Structure of an MLIP

Almost all MLIPs have the same qualitative structure, although the details of the implementation differ between MLIP types. The idea behind this structure is that for use in an MLIP, the local environment of all atoms must be represented by some set of numbers, or features, which we will call its atomic environment featurization (AEF). In **Figure 1B**, this is described as "local structural parameters". The AEF is built in a manner such that it can be represented as a manageable set of numbers, then that featurization is fed into an regression model. The accuracy of the model depends on how well these features and the model can capture the local environments, and, generally, larger sets of features are better able to capture environments (this is sometimes referred to as an AEF that is more "expressive").

## 4.2 Explicit vs. Implicit MLIPs

Determining how to distinctly categorize different MLIP approaches is challenging. This complication is the result of the multiple different ways researchers approach the featurization portion of MLIP development, and, as discussed below, how greater understanding in the field has prompted the convergence of various approaches, making the boundary between MLIP approaches more nebulous. However, we think that a helpful distinction at present is to consider MLIPs as being based on "explicit AEF" vs. "implicit AEF". We note that the designation of explicit vs. implicit AEF is analogous to what others, such as Schütt et al., have previously called "handcrafted" vs. "learned" representations.[39] By explicit AEF MLIPs, we mean MLIPs that define an explicit set of features for each element. Explicit AEFs are the type of potentials that were first invented by Behler and Parrinello[19] and have dominated MLIPs until quite recently, where the specific formulations of these explicit AEF MLIPs are discussed below in **Sec. 4.3**. In contrast, implicit AEF MLIPs are MLIPs that define a set of features or chemical descriptors which are learned, rather than pre-defined. Implicit AEFs result in learned features (sometimes called "embeddings") of the atoms and bonds comprising a material. MLIPs employing implicit AEFs will generally involve more ML architectural complexity, potentially making them harder or slower to use, train, and execute. Of particular importance is that, until recently,[54] the learned features from implicit AEFs can scale with number of different chemical species much more efficiently than those used in most explicit AEF MLIPs, and it is therefore this category of implicit AEFs that is almost always used for modeling many elements (e.g., > 5). As of this writing, implicit AEF MLIPs are almost entirely based on deep learning approaches for learning effective features. For example, implicit AEF MLIPs include all of the graph neural network (GNN) approaches (e.g., SchNet,[39] M3Gnet,[41] NequIP,[35] etc.) and the newest implementations of DeepMD.[21,22] Therefore, we will usually just refer to implicit AEF methods as deep learning methods, although these two categories are technically distinct.[39]In the text below, we will refer to explicit and implicit or deep learning-type MLIPs when the above distinction is useful.

### 4.3 Explicit AEF Type MLIPs

In this section, we describe the explicit construction of the AEF. The standard way to treat the mathematical representation of atom types and positions is to consider each atom as having an energy given by the atom type and its local environment (the positions and element types of nearby atoms). For this description, we refer to a given atom under consideration as the *target atom* (atom $i$ in **Figure 1**). The initial AEF for a target atom is generally constructed by writing the local atomic environment as a set of densities for a given atom type and then expanding that density function using a basis set consisting of radial and angular functions (for example, Bessel and spherical harmonic functions, respectively). The explicit AEF is most effective when it respects the symmetries of materials, which typically include permutations, translation, and rotation. A symmetry-aware representation can be created by taking tensor products of the initial AEF over the target atom and its near neighbors. These tensor products can then be combined to create a set of values that are covariant (i.e., change in a structured and predictable way) with symmetry operations. The final ML model then operates on these tensors, generally to predict a single scalar energy. It is possible and quite common to just keep scalar-covariant, generally called *invariant*, features, which can then be used in almost any ML model, provided the ML model is continuously differentiable.

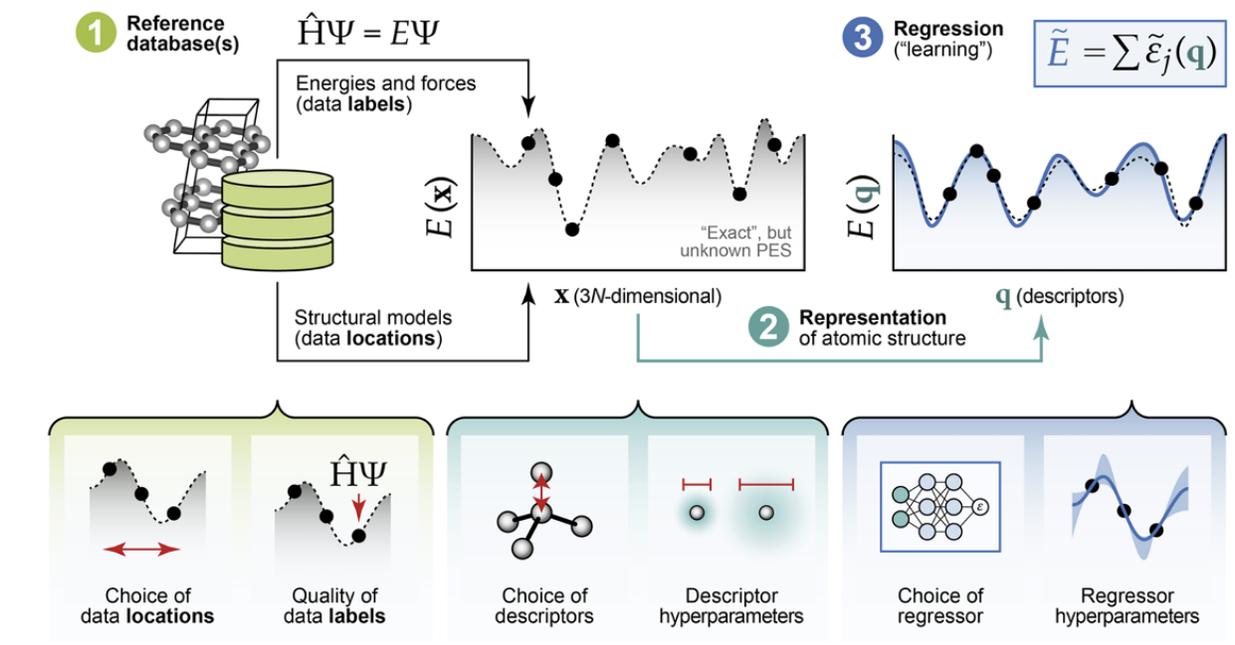

**Figure 2.** An overview of the explicit AEF approach of making an MLIP, including acquiring reference data from *ab initio* calculations, choosing a featurization approach to represent the local chemical environments, and an ML regression model to map the chemical environments to energies and forces. Adapted with permission from Ref. [50].

Once an AEF is established, any atom and its environment can be mapped onto an array, which can then be used as an input feature in a standard ML model. We call the simple passing of the AEF as features into the ML model the "explicit AEF approach". A graphical overview of the explicit AEF approach is given in **Figure 2**. The training target values for the ML model are typically a set of energies, forces and stresses. These targets could come from any source but are almost always taken from a large set of *ab initio* calculations, such as density functional theory (DFT). Any DFT cell calculation that provides target data, including stable structures, structures calculated during atomic relaxations, and structures calculated during AIMD are potentially of use. Then, the model parameters are estimated by standard regression methods. One difference from typical regression problems is that the training data are not just simple functions of the AEFs. First, the forces on a single target atom are often given as the derivative of the ML energy function with respect to the atomic position, and so in this case the fitting loss function must have a term that depends on the derivative of the ML energy function. It is worth noting that some MLIPs are trained on energies and forces separately, e.g., NN-based MLIPs trained without forces, where comparable accuracy can be obtained by increasing the training dataset size.[51,52] Second, each of the training energies are actually *total* energies for a whole set of atoms comprising a molecule or crystal unit cell (almost no *ab initio* methods allow unique formal decomposition of the total energy into values for each atom), and so the fitting loss function typically must have a term that depends on the sum of energies of all the atoms in each calculated configuration. Note that some MLIP code packages will fit to other properties as well, e.g., stress tensor, virial, polarizability, etc.[21,34] These additional properties can all be included in the fitting with regression approaches like those just described but with adjustment to the loss function. Assuming one is using an established MLIP code repository or package, these manipulations should be automatic and thus largely invisible to the user, and one can consider the MLIP fitting qualitatively as solving a simple regression problem. As with any regression, there are many possible ML models available. The most widely used models for MLIPs, listed in approximate order of their conceptual simplicity, decreasing speed of fitting and execution, and increasing accuracy are: Linear Regression (LR) (simplest, fastest, least accurate), Gaussian Process Regression (GPR), and Neural Networks (NNs) (complex, slowest, most accurate). We note that the speed of fitting GPR is highly dependent on the dataset size, and, for small datasets, GPR can have accuracies exceeding those of NN approaches, which tend to excel for problems involving large datasets.[1] Here, we exclude GNNs as they are discussed separately in **Sec. 4.4** in the context of implicit AEFs. From the above discussion, it is worth noting that some highly popular ML models used in regression problems, such as random forests, are not suitable for MLIPs because they do not possess continuous derivatives.

There is no universal answer to which ML model is best for MLIPs, but with good featurization, LR and GPR have both proven to work very well and are generally simpler to fit than NNs. Many of the most widely used MLIPs can be described with this explicit AEF framework. Specifically, the original Behler-Parrinello potential used atom-centered symmetry functions (ACSFs) as AEFs and a feed-forward NN ML model,[19] the Gaussian Approximation Potential (GAP) uses the Smooth Overlap of Atomic Positions (SOAP) approach to construct AEFs and a sparse-GPR ML model,[26] the Spectral Neighbor Analysis Potential (SNAP) used hyperspherical bispectrum functions (HBFs) as AEFs and a LR ML model,[40] the Moment Tensor Potentials (MTP) used moment tensor functions (MTFs) as AEFs and a LR ML model,[34] and the Atomic Cluster Expansion (ACE) uses the product of radial functions and spherical harmonics as its AEF and a LR ML model.[14] It should be noted that the ACE formalism was introduced as a superset of many other AEF methods, at least conceptually, meaning that ACSFs, SOAP, HBFs, and MTFs are all specific cases of ACE.[14] Note that this does not make potentials that use these earlier AEFs irrelevant, since any given potential may correspond to specific choices that are particularly efficient to train or execute, but it does help to realize that ACE appears to be a comprehensive formalism for expressing state-of-the-art explicit AEFs for MLIPs. Moreover, it is possible to combine any of these AEFs with any ML regression model. For example, the FitSNAP software[46] allows SNAP and ACE featurizations to be combined with PyTorch and JAX models.[53]

While the explicit AEF formalism is very effective, it has until recently had a significant scaling problem[13] which we describe here. We note this argument is based on the ACE basis construction, but it is quite general and similar issues occur in other related explicit AEF formalisms. Let $N_b$ be the number of basis functions used to expand the density of one chemical element around a target atom and $S$ be the number of elements. Then there are $N_{total\_basis} = N_b \times S$ total basis functions for one target atom. Let $v$ be the number of atomic sites we couple to in tensor products (here $v+1$ is called the body order, and $v+1 = 2$ gives pair information, $v+1 = 3$ gives 3-body information, and so on). For a given body order, there are order $O((N_b \times S)^v)$ basis functions. For a typical body order of 3, this gives quadratic scaling with the number of basis functions and species, which can become quite slow for a complex basis and for large numbers of elements. The element scaling typically limits most explicit AEF potentials to approximately 5 or fewer species.

### 4.4 Implicit AEF and GNN MLIPs

There are a few solutions to the issue of poor of scaling for explicit AEFs, particularly with the number of elements. The general approach to overcoming this scaling problem is to instead use an implicit AEF to embed the chemical space in a learned feature vector (i.e., an embedding) that can effectively represent different chemistries without explicitly developing basis functions for each one. This approach appears to work very well, dramatically reducing the complexity of treating different elements. The exact reason this works is not totally clear, but likely is because the properties of different elements are not independent, and their interactions in subclusters inform their more complex cluster interactions (e.g., pair couplings can dominate the energy of a cluster of 10 different atom types). Probably the most widely used approach that provides efficient embedding (as well as has other potential advantages and disadvantages) are GNNs, discussed more below. However, there are other approaches. For example, the DeepMD[21,22] MLIP represents the local environment as embedding vectors that are constructed by a neural network based on some or all of local atom distances, angles, and types. The weights of the embedding network are trained during fitting, making the AEF an implicit function of the coordinates that is learned during training and allowing DeepMD to fit many elements. A number of papers have recently shown how features in standard explicit AEF MLIPs might be manipulated to reduce the scaling with species, where such an approach by Lopanitsyna, et al. is illustratively named "chemical compression".[54–56] Darby et al. in particular has shown that linear embedding of the elements into a fixed dimensional vector space corresponds formally to tensor-decomposition, and as the dimension increases will converge to the uncompressed result.[54] Artrith et al. showed that element-specific weights allow constant-size AEF vectors irrespective of the number of chemical elements and demonstrated the method for up to 11 species[16] and a similar approach was independently proposed by Gastegger et al.[57] An outstanding example of the power of these approaches is the MACE and the graph ACE (grACE) packages (corresponding to independent implementations of the same underlying architecture), both of which provide excellent scaling with the number of elements while achieving high accuracy.[29] These recent papers and emerging packages suggest that soon the chemical scaling issues associated with the explicit AEF approach may be greatly reduced or removed altogether.

A GNN is an NN architecture that operates on graphs, where graphs are collections of nodes and the connections between them (called edges). Perhaps not surprisingly, a graph is an excellent way to think about interacting atoms, where nodes are mapped to atoms and edges are mapped to all reasonably near neighbor bonds (but usually much longer than covalent bonds, e.g. 4-6 Å). So, the "graph" in this case essentially serves to encode the localization of forces and other properties. One might argue that this same local structure is also encoded in the finite cutoffs in all AEFs mentioned above and used in all MLIPs. But GNNs and the explicit graph representation lead to a somewhat different approach to constructing AEFs with some clear advantages vs. the explicit AEF approaches discussed above, and therefore have become a very popular approach for MLIPs. In the graph, sets of embeddings can be freely associated with both nodes and/or edges, and these embeddings can be mapped to properties of the atoms by the GNN. GNNs iteratively update the embeddings of a target node/bond through learned mappings of connected node/bond embeddings onto the target node/bond, with the connections determined by the graph structure.

Each one of the updates is typically done in one layer of the GNN. These updates are also given structural information like bond lengths or more detailed AEF parametrizations. Because GNNs encode the features of atoms and bonds through a learned mapping to embedded features, these features can potentially represent the chemistry and structure much more efficiently than the basis function tensor products in the standard explicit AEF MLIP described in **Sec. 4.3.** There is much freedom in constructing GNNs. For example, recent GNNs are often so-called E(3) equivariant NNs (e.g., NequIP[35][41], MACE[32], TeaNet[58]), which work with highly expressive equivariant tensor representations of atomic environments and operate on them to preserve the proper symmetries. Such GNNs appear to be particularly data-efficient in fitting. A systematic overview of the design choices has recently been published.[59] Most GNN MLIPs effectively couple a widening range of atoms/bonds to a target atom/bond at each layer of the GNN. When many layers are needed to get good convergence they end up effectively coupling atoms that are 3-4 nm apart. This coupling can be advantageous for capturing longer range interactions, e.g., as shown for M3GNet in comparison with MTP potentials.[41] However, this medium range coupling of 3-4 nm is much longer than typical ranges of direct physical interaction in almost all PBPs and explicit AEF MLIPs (which are almost always 1 nm or less) and can lead to significant memory and parallelization issues. Therefore, researchers are now exploring more local equivariant NN approaches, e.g., Allegro,[13] that forgoes the message passing structure on the graph altogether, or MACE, [31] which uses high body order local features such that only two hops on the graph are sufficient for convergence. Both show excellent scalability with multiple processors.

## 4.5 Unifying explicit and implicit AEFs

It is worth noting that all of these MLIP methods are increasingly appearing to be different aspects of a single general MLIP approach. As discussed above, explicit and implicit AEFs were developed largely independently. Explicit AEFs focus on local descriptions of atomic energy obtained by the interaction with all neighbors within a cutoff distance. Implicit AEFs recursively incorporate via message passing information about atoms that can be several cutoff distances away. The messages are assembled from the local atomic environment within a cutoff distance and then employed for the computation of the energy of another atom. From the viewpoint of explicit

AEFs, message passing modifies the character of an atom. In an explicit AEF neighboring atoms are characterized by their positions and chemical elements. In an implicit AEF neighboring atoms are characterized by further attributes collected from the atomic environment. For example, this makes a carbon atom on a surface different from a carbon atom in the bulk. In equivariant neural networks the additional attributes are vectors and tensors, which essentially give the carbon atom an environmentally dependent, non-spherical character.

ACE provides a complete basis for the local atomic environment. Applied to the local atomic environment of neighboring atoms, ACE facilitates formally complete messages.[59] Recursive application of ACE in an implicit AEF construction leads to "multi" ACE,[60] and also MACE[31]

However, while intuitive, it is not necessary to take an iterative evaluation as the starting point. One can get to the same model by initially considering the most general graph basis functions, as in the grACE method.[29] In this setting, the ACE basis can be seen as using only star graphs. In complete analogy to the original ACE model, the energy or any other local or semilocal property is written as a linear combination of general basis functions defined on the graph, i.e. looking like an explicit AEF. However, the cost of enumerating general basis functions on graphs formally scales exponentially with the number of nodes and/or edges. The efficient evaluation employing tensor decomposition along the graph ACE basis functions, is iterative evaluation that corresponds to tree graphs and is equivalent to message passing neural networks such as NequIP,[35] MACE, i.e. we recover to implicit AEFs.

This facilitates the following understanding. Explicit AEFs on graphs that have tree structure with two or more layers can be transformed to implicit AEFs for numerically efficient evaluation, resulting in message passing neural networks. In practice even single layer explicit AEFs are evaluated iteratively in terms of body order for numerical efficiency,[61] which suggests that they could also be thought of as an implicit AEF.

Therefore, the distinction between explicit and implicit AEFs reflects the history of the development of AEFs in the past years, and their implementation details, more than their actual underlying mathematical structure. To the best of our knowledge, all AEFs can be represented in an explicit way. Iterative evaluation for numerical efficiency leads to implicit representations of AEFs. These results increasingly suggest that we may be converging on a single general formalism for MLIPs, and the seemingly very different approaches in use today are actually specific choices within the general design space. Such understanding will hopefully allow the community to extract the approaches that are simultaneously optimized to be the most efficient for training, and fast and most accurate for prediction.

## 5    Universal MLIPs

To date, the vast majority of MLIPs are trained on a limited domain of chemical or materials systems, specific to a given application interest. This amounts to an MLIP that represents a particular materials family (e.g., perovskite oxides, 2D MXenes, etc.) or particular chemical system (e.g., the Li-Co-Mn-Ni-O composition space) well, but is not transferable in the sense that these MLIPs are unable to extrapolate to accurately model new elements or local structures that are not present in the specific training data. As discussed in **Sec. 4** above, part of the reason researchers focus on small numbers of chemical elements is related to the explicit AEF approach used, which in the past did not scale well to many elements. The creation of accurate, highly general MLIPs that cover many more elements and conditions than typical MLIPs is highly desirable as it would produce a potential with the widest possible domain of applicability, enabling the study of the statics and dynamics of many types of chemically complex systems, potentially for long simulation time scales. In thinking about scaling up MLIPs to more chemical species, we propose that it is useful to distinguish a few categories of MLIPs, specifically:

1. Targeted MLIPs (T-MLIPs). These are the typical MLIPs that cover approximately 1-10 (usually $< 5$) elements and are typically under some constraints of chemistry, structure, or phase (e.g., oxides with certain compositions, multiple elements in a fixed crystal structure for high entropy alloys, or molten (liquid) phase salts) although these latter constraints can be quite few or potentially even none.

2. Universal MLIPs (U-MLIPs). These attempt to cover a large number of species under different levels of constraints, e.g., transition metal oxides in solid form or organic molecules with select heavy elements. These typically cover 10-100 elements and could range in conditions, from a very strong constraint (e.g. a specific crystal lattice) to allowing almost any atomic configuration. Obviously, the MLIP would be considered more useful and universal as more elements are included and fewer constraints on the considered chemical or material structures are made. It can be useful to consider these MLIPs in two categories, which we call semi-universal-MLIPs (SU-MLIPs) and true U-MLIPs. Both require a method that can scale well with number of species and target a large number of species. However, SU-MLIPs focus on a select domain, e.g., transition metal oxides in solid form or organic molecules with select heavy elements. A good example of an SU-MLIP is the recent AIMNet2,[18] which targets molecular and macromolecular structures and is applicable to species containing up to 14 chemical elements in both neutral and charged states, making it valuable for modeling the majority of non-metallic compounds. As another example of a SU-MLIP, the work of Rodriguez et al. built the Elemental Spatial Density Neural Network Force Field (Elemental-SDNNFF), which produces accurate forces for Heusler alloys constituting 55 different elements and accurate predictions of phonon properties.[23] A third example is the SuperSalt potential from Chen et al., which models M-Cl molten salts for 11 cations M, and was shown to be significantly more accurate than existing U-MLIPs for these materials. In contrast, U-MLIPs attempt to cover a very large fraction or even almost all of the periodic table with atoms potentially in any arrangement. Even for U-MLIPs, it is typical to exclude elements that are very impractical or intractable to *ab initio* methods, e.g., Nobelium (atomic number 102 or anything with an atomic number above 103). Thus, the relevant portion of the periodic table for materials and chemistry is generally up to about 100 elements. U-MLIPs typically cover over 50 elements and may accurately model solids, liquids, and molecular structures. A good example of this class is the recent M3GNet potential from Chen and Ong,[41] with 89 elements and no particular constraints on its applicability (although there is a strong bias in training to solid phases), or the above-mentioned MACE-MP0, which is trained on the same data and was shown to be effective for running stable MD simulations for many nanoseconds.[32]

The exact values of the number of elements or level of structural constraint in the categories above are somewhat arbitrary, although T-MLIPs are distinct from U-MLIPs in that the latter typically require scalable implicit AEF methods (see **Sec. 4**). In particular, in this section U-MLIPs will be used rather loosely to indicate an MLIP which has been trained on sufficiently large and diverse datasets such that it provides usefully accurate predictions on a wide range of compositions and structures for molecules and/or materials. If the training data is sufficiently large and diverse, the MLIP may provide accurate predictions for the behavior of most chemically relevant elements in the periodic table.

Universal potentials are not limited to MLIPs and have been developed previously in the context of PBPs. The creation of semi-universal PBPs (SU-PBPs) dates back to 1981 with the seminal work of Weiner et al.[62] Since this time, the semi-universal force field (UFF) of Rappe et al.[63] and the Assisted Model Building with Energy Refinement (AMBER) force fields[62,64] have emerged as some of the most popular SU-PBPs, where the main utility of these potentials is for modeling molecular systems (e.g., to aid drug discovery), as opposed to condensed phases. The relative utility of these SU-PBPs vs. U-MLIPs is difficult to determine at this stage since the development of U-MLIPs is still in the nascent stages. The first reported SU-MLIP for organic systems (representing molecules initially containing only C, H, O, N atoms) is the Accurate NeurAl networK engINe for Molecular Energies (ANAKIN-ME, ANI for short) potential from the work of Smith et al. in 2017,[12,65] which was expanded in 2020 to include S, Cl and F elements (thus covering ~90% of drug-like molecules).[66] The ANI potential has similar applicability as the universal AMBER PBP for organic systems, but in MLIP form. The latest iteration of this SU-MLIP as of this writing came in late 2023, termed the atoms-in-molecules neural network potential (AIMNet2) SU-MLIP.[18] This SU-MLIP extends the ANI potential to include up to 14 elemental species and additional energy terms related to short-range van der Waals (vdW) correction and long-range electrostatic correction, enabling higher fidelity predictions of organic molecules and macromolecules which can also include the effects of charged species and species with different valence states. In addition, the ANI-1xnr potential extended the success of the ANI SU-MLIP to also enable the accurate modeling of condensed phases of organic systems (comprising C, N, H, O) such as liquids, supercritical fluids, and chemical reactions.[67] Finally, the MACE-OFF3 potential,[33] also published in late 2023, uses the MACE message-passing framework to construct a SU-MLIP for the 10 most-occurring elements in organic chemistry (H, C, N, O, F, P, S, Cl, Br, I). Compared to the most recent ANI potentials, MACE-OFF3 uses only short-range interactions, yet results in improved performance on a number of benchmark molecular simulation properties compared to ANI.

The first published U-MLIPs intended to have broad applicability across most elements in the periodic table came nearly simultaneously in early 2022.[41,58] In just the past two years, many U-MLIPs capable of representing most elements in the periodic table have been developed: (1) the 3-body Materials Graph NETwork (M3GNet) potential from Chen and Ong;[41] (2) the Crystal Hamiltonian Graph Neural Network (CHGNet) of Deng et al.;[20] (3) the unified atomistic line graph neural network-based force field (ALIGNN-FF) of Choudhary et al.;[17] (4) the tensorial message passing neural network PreFerred Potential (PFP) from the work of Takamoto et al.,[37] which is now shared as a commercial product in the Matlantis package;[68] (5) the Graph Networks for Materials Exploration (GNoME) U-MLIP from Merchant et al.,[28] which is a custom-trained version of NequIP from the work of Batzner et al.[35] fit to an in-house database of roughly 80 million DFT calculations;[28] (6) the SevenNet-0 potential from Park et al.,[38] which is also based on NequIP and trained in the same Materials Project data as M3GNet but refined to provide good scaling on many processors for modeling larger systems; (7) the equivariant graph tensor network MACE-MP0, developed by Batatia et al.,[32] which was trained on the same publicly available data used by the CHGNet and SevenNet models, and was demonstrated to have a high degree of accuracy on three illustrative applications of dynamics of aqueous systems, heterogeneous catalysis, and metal-organic frameworks but also showed stable nanosecond-long molecular dynamics on over 30 examples with diverse chemistry; (8) the graph-based pre-trained transformer force field (GPTFF) developed by Xie et al.,[27] a GNN model with transformer blocks integrated into the model architecture; (9) MatterSim,[30] a large-scale deep learning model from researchers at Microsoft trained on actively-learned DFT data from a large custom database of roughly 17 million atomic configurations, including many non-ground state structures over a large temperature (0-5000 K) and pressure (0-1000 GPa) range; (10) the Orb model developed by Neumann et al.[36] achieves excellent performance on the MatBench leaderboard and offers the advantage of faster performance compared to other leading U-MLIPs, where, for example, it was found Orb performed 3-6 times faster than the pytorch version of MACE (but with advantage disappearing when custom CUDA kernels are used), particularly for large system sizes and if dispersion corrections were included; and, finally, (11) the EquiformerV2-OMAT24 model from Meta,[24] which trains the EquiformerV2 model[69] on a novel open source database of roughly 118 million atomic configurations, leading to, as of this writing, the best performance on the MatBench leaderboard. An overview of some example capabilities of U-MLIPs is given in **Figure 3**. The main ranking of the MatBench leaderboard is by the ability to predict stability of unseen materials. However, prediction of properties is quite a bit more difficult, and the recently added data showing the ability of U-MLIPs to predict thermal conductivity shows little correlation with the main ranking, with a MACE model outperforming all other models by quite some margin. This highlights the danger in using only results sensitive to energy predictions to extrapolate to general performance. A note of caution: the field is very active, and while the comments herein represent the status as of writing, it is likely that new advances in performance will keep coming in the near future.

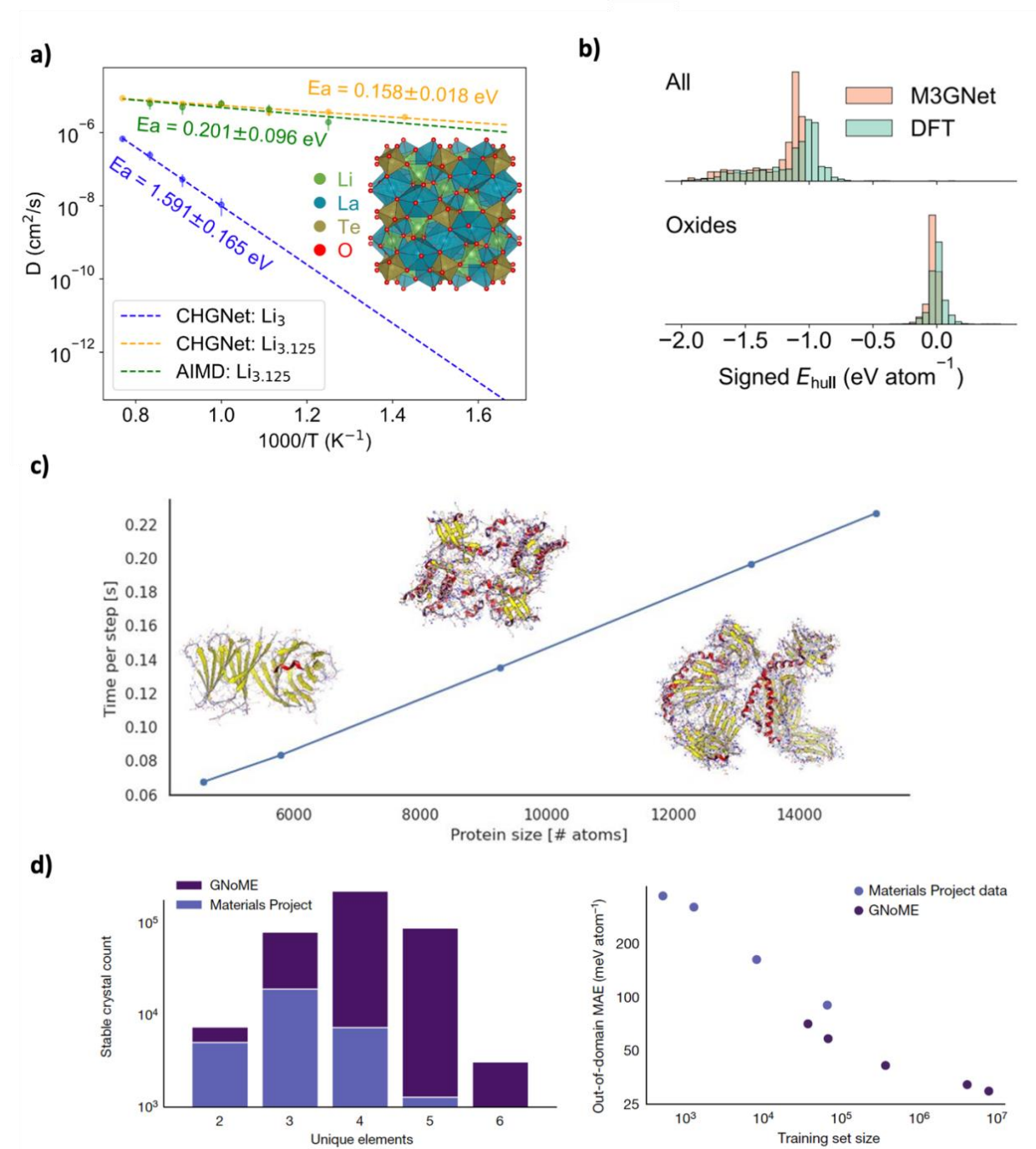

**Figure 3.** Examples of applications of U-MLIPs. (a) Li diffusivity in the solid electrolyte material $Li_3La_3Te_2O_{12}$ using the CHGNet potential, adapted with permission from Ref. [20]. (b) Calculated thermodynamic stabilities (as signed convex hull distance) of a set of hypothetical predicted

materials using the M3GNet potential and compared to DFT calculations, adapted with permission from Ref. [41]. (c) Time to optimize the structure of large protein structures using AIMNet2, adapted with permission from Ref. [18]. (d- left) Number of newly discovered stable materials using a GNN model that only predicts the formation energy of a given crystal based on number of unique elements in the structure, (d- right) mean absolute error as a function of training set size using the same GNN, adapted from Ref. [28].

The above U-MLIPs are made possible by advancements in previously developed GNN models to include physical information of how the bond energies of a system evolve with the positions of the constituent atoms, enabling the acquisition of forces and stresses via differentiation of this learned energy dependence. For example, the M3GNet potential is an extension of the MatErials Graph Network (MEGNet) model[70] to include 3-body interactions (note general N-body interactions are possible, but 3-body is used for computational efficiency), explicit atomic coordinates, and the 3×3 crystal lattice matrix.[41] As another example, ALIGNN-FF extends the ALIGNN model,[71] which already incorporates many-body interactions, to also produce atomwise and gradient predictions, thus enabling calculation of the force on each atom and stress on the system.[17] In addition to advancements to underlying GNN models, U-MLIPs have been made possible by the growth of large computational databases, namely those containing millions of static DFT calculations and AIMD simulations. Each cataloged DFT structure provides one energy and 3N forces (N = number of atoms in the structure) to use for training the universal MLIP. The presently available U-MLIPs were all trained on various databases of DFT calculations, as summarized in **Table 1**. In addition, **Figure 4** shows the evolution of DFT database size used to train various U-MLIPs over time. We find, on average, that the database size has increased by more than an order of magnitude each year, from roughly $2 \times 10^5$ in 2022 (M3GNet) to a present maximum of $1.18 \times 10^8$ in 2024 (EquiformerV2-OMAT24). Even one more year of following this trend would bring the community to the level of training on billion calculation databases, a demanding goal but one that would likely bring further improvements in performance.

**Table 1.** Summary of data and applicability domain of U-MLIPs.

| U-MLIP name | Training database | Number of elements represented | Training data amount | Notes |
|---|---|---|---|---|
| M3GNet | Materials Project | 89 | 62,783 compounds: 187,687 energies, 16,875,138 forces, | Training data taken from Materials Project dating back |

| | | | and 1,689,183 stresses | to its inception in 2011 |
|---|---|---|---|---|
| CHGNet | Materials Project + Trajectory database | 89 | 146,000 compounds: 1,580,395 energies, 49,295,660 forces, and 14,223,555 stresses | Training data taken from Materials Project GGA and GGA+U relaxation trajectory up to Sept 2022 version. |
| ALIGNN-FF | JARVIS-DFT | 89 | 307,113 energies and 3,197,795 forces for 72,708 compounds | |
| PFP (Matlantis) | Custom | 96 (previous versions were for 18 (TeaNet) and then 45 elements) | Roughly 10 million configurations | Training data is a custom in-house set performed by a collaboration of Preferred Networks, Inc. and the ENEOS Corporation |
| GNoME | Materials Project + Custom | 94 | Roughly 89 million configurations from 6 million compositions | Initial training done on Materials Project data from 2018 comprising 69,000 materials. Later fits include about 89 million configurations |
| MACE-MP0 | Materials Project + Trajectory database | 89 | Same training data as used to build the CHGNet potential | An additional dispersion correction model can be used to accurately capture dispersion physics not present in the training data |
| SevenNet-0 | Materials Project | 89 | Same training data as used to build the CHGNet potential | Same training data as used to build the M3GNet potential |
| GPTFF | Atomly.net | Value not given in text | Roughly 2.2 million crystal structures, consisting of a total of 37.8 million energies (349k of these are equilibrium states), 11.7 billion force vectors, and 340.2 million stresses | |
| MatterSim | Initial data from public databases like Materials Project, Materials Project Trajectory, and Alexandria, then customized with | 89 | Roughly 17 million atomic configurations | Sampling techniques include simulations with temperatures ranging from 0-5000 K and pressures from 0-1000 GPa |

|  | additional DFT calculations |  |  |  |
| --- | --- | --- | --- | --- |
| Orb | Materials Project Trajectory and Alexandria | 89 | Value not directly mentioned in text |  |
| EquiformerV2-OMAT24 | Initial data from public databases like Materials Project, Materials Project Trajectory, and Alexandria, then customized with additional DFT calculations | 89 | Roughly 118 million atomic configurations | As of this writing, state-of-the-art performance on the main ranking of the MatBench leaderboard and largest publicly-available DFT database |

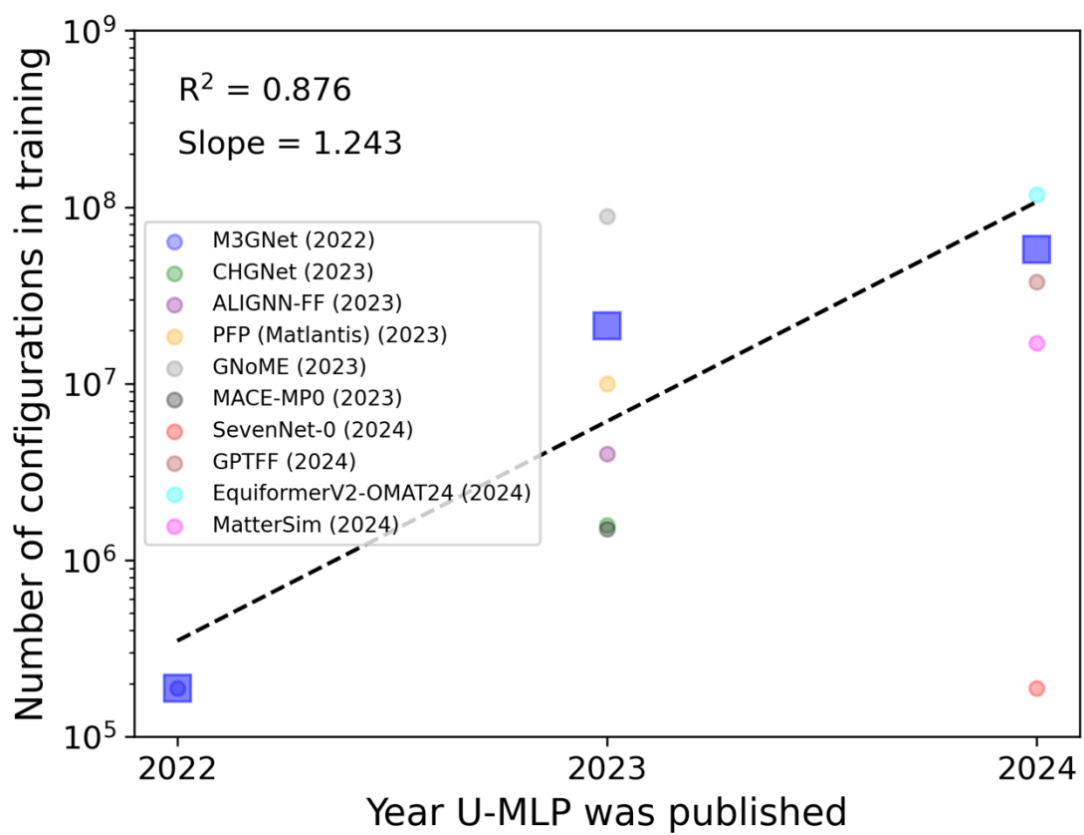

**Figure 4.** Evolution of DFT database size used to train U-MLIPs over time. The small circle points are values for individual U-MLIPs, and the large blue squares are the average for a given year (note that SevenNet-0 was not included in the average for 2024).

While it is possible to train a U-MLIP on only energies, Chen and Ong recommend training on energies, forces, and stresses to obtain the most physically accurate potential, and the inclusion of stresses is needed if one is interested in modeling structural phase transformations or performing molecular dynamics simulations where volume can vary (e.g., NPT ensemble).[41] These U-MLIPs tend to have very good accuracy when averaged over large test data sets, evidenced by test errors in CHGNet (M3GNet) which has energy, force and stress mean absolute errors on test data of 29 (35) meV/atom, 70 (72) meV/Angstrom, and 0.308 (0.41) GPa, respectively. ALIGNN-FF, trained only on energy and forces as seen in **Table 1**, has energy and force mean absolute errors on test data of 86 meV/atom and 47 meV/Angstrom, respectively. Note the higher errors for ALIGNN-FF are likely due to the authors using roughly 300k of the 4 million data points available to them for training due to hardware and speed constraints, and likely not a fundamental limitation of the ALIGNN-FF approach.

In general, the developers of these U-MLIPs (M3GNet, CHGNet, ALIGNN-FF, PFP, GNoME, MACE-MP0) all perform multiple benchmark tests on various classes of materials structures, chemistries, and prediction of resulting materials properties. While the specific tests and comparisons are too numerous to list here and also not directly comparable due to different databases used for training and testing, all of these U-MLIPs are successful in accurately modeling a very large domain of materials phenomena, with typical energy, force and stress errors greatly surpassing many-body PBPs such as EAM and modified EAM and achieving comparable or slightly worse accuracy than explicit AEF approaches relying on local environment representations like MTP. Therefore, it appears possible these U-MLIPs may soon be able to achieve near DFT accuracy across many different arrangements of atoms.

The CHGNet U-MLIP is unique from the other U-MLIPs discussed here because it additionally includes the electronic effects of valences by explicitly embedding the magnetic moments on the vector representation of each atom, thereby enabling charge-informed atomistic simulations.[20] The inclusion of such electronic effects in an MLIP might be beneficial to modeling some materials phenomena that are highly correlated with charge states (i.e., transition metal bonding dictated by the ions' valence states, and phase transformations driven by charge disproportionation, discussed more below). There are different approaches to represent charge on an atom, and in CHGNet, the charge is inferred via the DFT-calculated magnetic moment, which is essentially the localized spin density that is governed by the electron orbital occupancies of a given valence. Therefore, the training data and predicted outputs of CHGNet consist of energies, forces, stresses, and magnetic moments on every atom in the system, where the addition of magnetic moments in training led to further error reductions of energy, force and stress (in the range of 1-10%, depending on the property) compared to not including magnetic moments in training. More important than slight error reductions is the new ability to model key pieces of physics governed by specific valence states and charge transfer which was not possible with any previously formulated MLIP. To illustrate the power of this capability, Deng et al.[20] highlight the ability of CHGNet to (i) accurately discriminate different valence states of transition metal with the example of V oxidation in $Na_4V_2(PO_4)_3$, (ii) enable the study of charge transfer-based dynamic information with the example of charge-coupled degradation in $LiMnO_2$ battery cathode material, where the degradation is driven by the dynamic differences of $Mn^{2+}$ and $Mn^{3+}$ vs. the immobile $Mn^{4+}$ cations, and (iii) model how the electronic entropy effects in the battery cathode material $Li_xFePO_4$ drives the finite temperature phase stability of this material, where the inclusion of Fe valences in CHGNet correctly reproduces the qualitative miscibility gap as Li is added to $Li_xFePO_4$, whereas no miscibility gap is observed if the Fe valence effects are ignored. Finally, it is worth noting that while the original CHGNet model took 8.3 days to train on a single A100 GPU, the recently developed FastCHGNet includes several optimizations which results in significantly faster training, down to just 1.5 hours when using 32 GPUs.[72]

There have been at least five notable, recent studies benchmarking the performance of different U-MLIPs. First, work by Yu et al.[73] compared the ability of M3GNet (and the newer Pytorch-based MAT-GL implementation), CHGNet, MACE-MP0, and ALIGNN-FF to predict various materials properties. Regarding the convergence behavior of cell relaxations, they found CHGNet and MACE-MP0 to be best, with M3GNet having numerous cases of providing non-converged full-volume relaxations. All models could predict formation energies roughly as well, though CHGNet had the lowest MAE at just 81 meV/atom, while all other models had MAEs that were 129 meV/atom or higher. For vibrational properties, MACE-MP0 emerged as the best, while ALIGNN-FF demonstrated some significant qualitative errors with reproducing phonon band structures. Second, work by Focassio et al.[74] compares predictions of M3GNet, CHGNet, and MACE-MP0 for predictions of bulk and surface total energies and surface energies for 73 elemental systems for which bulk and surface slab data were available in the Materials Project, where a total of 1497 surface structures were considered. As shown in **Figure 5**, all three U-MLIP models were able to accurately reproduce the total energies of bulk (note, on average CHGNet has the lowest prediction errors), which is sensible as these bulk structures were included in the U-MLIP training data. The errors for surface energies are much more significant than for bulk, which is the result of these surfaces not being present in the training data. Surface energy prediction errors with M3GNet and CHGNet show systematic underprediction and MACE-MP0 shows multiple instances of overprediction. Focassio et al. also show that targeted MLIPs like MTP and NequIP can have lower errors for predicting properties of specific systems vs. the zero-shot U-MLIP predictions, improving accuracy at the cost of losing generality. The third benchmark from Deng et al.[75] shows the underprediction of energy and forces by U-MLIP in a series of material modeling tasks, including surface energy, defect energy, mixing energy, phonon vibrations, ion migration barriers, etc. The observation of underpredicted properties aligns with the report by Focassio et al. The underpredicted energies and forces are attributed to a systematic softening of the U-MLIP PES, where the U-MLIPs are found to predict smoother energy landscapes than the real PES described by DFT. The author claimed the softening effect is driven by the biased sampling in U-MLIP training dataset, where the training atomic configurations are taken from DFT ionic relaxations and are therefore close to local PES minima. The fourth benchmarking work we discuss here is from Riebesell et al.,[76] who focused on the ability of U-MLIPs and other GNN-based ML models (e.g., MEGNet, ALIGNN) to predict stable materials (i.e., materials with a convex hull energy within some threshold, chosen as being on or below the Materials Project training data convex hull). They tested these models on the dataset from Wang et al.,[77] which consists of unrelaxed structures of materials less well-sampled in the Materials Project and was generated by a chemical-similarity based element substitution process using structures from the Materials Project. They found that all three U-MLIP models outperformed all other models, and that, in particular, MACE-MP0 performed best for discovering new stable materials, where the classification F1 scores for finding stable materials followed the order of 0.67 (MACE-MP0) > 0.61 (CHGNet) > 0.57 (M3GNet) > (everything else). The MACE-MP0 and CHGNet models had MAE values of convex hull energy of 60 meV/atom. Finally, work from Casillas-Trujillo et al. sought to evaluate the ability of M3GNet, CHGNet and MACE-MP0 to predict metallic alloy mixing thermodynamics. A striking result of their work is that none of these 3 U-MLIPs were able to accurately reproduce the mixing energies of metallic binary alloys in adequate agreement with DFT results.[78] These findings point to the need for careful benchmarking when pursuing the use of U-MLIPs for a new problem of interest, and, if sufficient accuracy is not obtained, the consideration of carefully selected additional training data to fine-tune the U-MLIP to obtain enhanced accuracy. To this end, recent work from Wines and Choudhary established the Computational High-Performance Infrastructure for Predictive Simulation-based Force Fields (CHIPS-FF), which is an open-source infrastructure specifically tailored for benchmarking materials properties predicted with various U-MLIPs.[79] The problems identified in the above benchmarking studies suggest a benefit to a focusing on not just force and energy errors but also quantitative assessment of U-MLIP errors on physically relevant properties, e.g., surface energies, defect energies, mixing behavior, elastic constants, etc. Testing of such behavior will benefit from the careful generation of sophisticated test datasets to assess U-MLIP performance. There are challenges for how to do this effectively since, once a test set is established, it is tempting for the community to begin effectively fitting new potentials to minimize errors on these test sets, which may inadvertently create undesirable performance of the potential with respect to other properties. Developing and properly utilizing such test datasets is expected to play an important role in the refinement of U-MLIPs. It is important to note the that the limited ability for low energy and force errors on training and test data to assure good performance in predicting important materials properties is not limited to U-MLIPs and is a challenge for MLIPs. We discuss this issue further in **Sec. 7.5.3**.

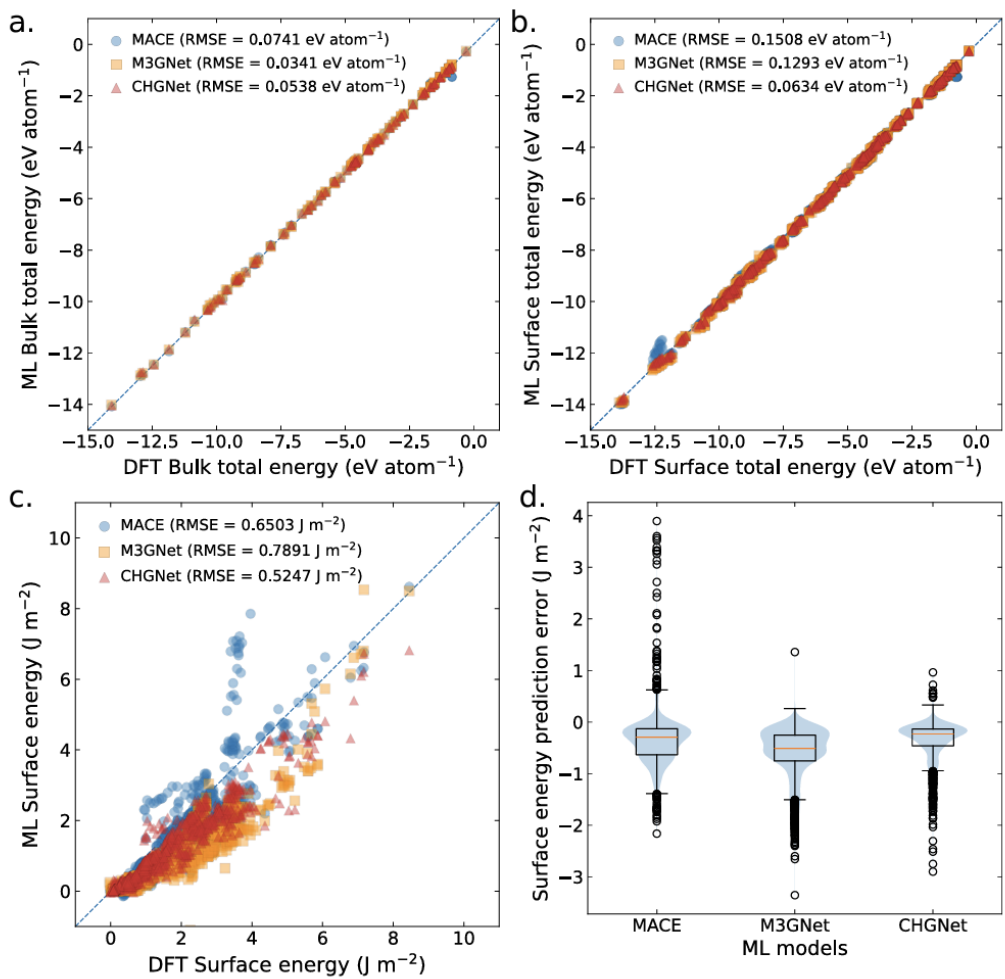

**Figure 5.** Benchmarking performance of U-MLIP models for predicting bulk and surface total energies and surface energies. ML vs. DFT total energies for (a) bulk and (b) surface. (c) ML predicted vs. DFT-calculated surface energies. (d) Data from (c) but plotted as a box-and-whisker plot of the ML vs. DFT residuals. Figure adapted with permission from Ref. [74]. MACE refers to the MACE-MP0 model.

As mentioned above U-MLIPs are expected to keep improving rapidly. Such improvements can come from simply refitting a potential to more data e.g., as already demonstrated by Takamoto et al.[37,58,80] Improvement can also come from expanding the underlying MLIP formalism to include new pieces of physics, as was done by Anstine et al. to include vdW and electrostatic contributions to the total energy of the organic U-MLIP AIMNet2,[18] and the addition of dispersion and vdW interactions to MACE-MP0 despite the potential only being trained on PBE-level DFT data.[32] A different and more subtle method of using additional data to improve a U-MLIP is through fine-tuning of an existing model. Fine tuning is a process by which a large NN model that is already trained has its weights only slightly altered to match a small amount of new data, with the goal to keep the model accuracy on its original training data while increasing the model accuracy on the new data. Fine-tuning typically involves updating a small fraction of the weights, typically in layers involved in just the final steps before output. Such fine-tuning has been widely applied in other ML problems (e.g., computer vision and language models). Through this approach, U-MLIPs may form the basis for more focused models that can be fine-tuned using new data comprising more specific chemical or structural families of materials or molecules. Since the weights in the U-MLIPs have been pre-conditioned on comprehensive datasets, the fine-tuning process typically requires fewer data compared to fitting a new potential, and may result in lower errors than training from scratch.[28,75,81] For example, Merchant et al. found that the error of a fine-tuned U-MLIP also follows a power-law as a function of its pretraining data size[28] (i.e. larger pre-training dataset sizes led to better downstream fine-tuned U-MLIP's. The M3GNet, CHGNet and MACE Python packages already allow for fine-tuning, so this approach can be readily explored by users.

    U-MLIPs have several promising use cases. The first is that they may drastically speed up DFT calculations by providing a means to quickly relax a set of atomic positions much closer to equilibrium positions prior to running a full DFT calculation. In their work on developing M3GNet, Chen and Ong discuss how such speedup may reduce DFT calculation time for relaxing material structures by a factor of three.[41] This application is largely insensitive to inaccuracies in the U-MLIP since the final output is from a full *ab initio* calculation and it is therefore extremely appealing. One could imagine it becoming standard practice and having a large impact, cutting typical *ab initio* calculation times significantly across potentially billions of future calculations. A second use case is replacing and expanding beyond AIMD. Similar to all MLIPs, U-MLIPs are useful for simulating large-scale, long-time dynamic phenomena inaccessible to current AIMD length and time scales. Such speed-up of DFT and MD studies has the potential for disruptive transformation of atomistic modeling, potentially impacting thousands of studies each year. A third use case is materials exploration. Different from more targeted MLIPs, U-MLIPs have a much broader domain of applicability, increasing the chemical and structural complexity of systems that can be modeled with typically a small or minimal loss in accuracy. This makes U-MLIPs particularly powerful for exploring many chemistries and structures, e.g., screening for a certain property like Li-ion conductivity or high elastic modulus. In particular, the lack of scaling issues with many components makes U-MLIPs uniquely positioned for exploration of chemically and structurally complex multicomponent materials. There is thus a large opportunity to screen materials properties across the periodic table which was only possible with computationally expensive *ab initio* calculations in the past, but which could be made roughly 1000× faster for even modest-size unit cells with the aid of U-MLIPs. As a demonstration of the beginnings of such an approach, Chen and Ong developed matterverse.ai, a Materials Project-like repository containing millions of hypothetical structures generated using physics-based considerations of reasonable materials structures and chemistries, and for which formation energies were subsequently calculated and screened using the M3GNet potential.[41] Similarly, Merchant et al. used a GNN model that directly predicts the formation energy of a crystal to propose 381,000 new stable (at T = 0 K) materials, expanding the number of inorganic materials predicted to be stable by nearly an order of magnitude. Given the rapid advances in generative AI, one can imagine the possibilities of combining generative inverse materials design approaches together with U-MLIPs for fast materials exploration and screening, for example using tools like the Crystal Diffusion Variational Autoencoder (CDVAE) of Xie et al.[82,83], the MatterGen model of Zeni et al.[84], the Symmetry-aware Hierarchical Architecture for Flow-based Traversal (SHAFT) model of Nguyen et al.[85], a diffusion probabilistic model employing unified crystal representations of materials (UniMat) from Yang et al.,[86] or even using large language models trained to produce stable crystal structures,[87,88]. Joining generative and U-MLIP methods may provide a powerful new way to discover exceptional new materials that would not have been considered by way of conventional screening approaches.

Presently, the main drawbacks of U-MLIPs include the same limitations as noted for more targeted MLIPs (see **Sec. 9**) with the additional limitation that its true domain of applicability is quite uncertain. While U-MLIPs are much broader in their domain than targeted MLIPs, the currently available models almost certainly have many areas of major weakness that cannot be easily predicted in advance. For example, using a U-MLIP to study Li diffusion in solid electrolytes might provide excellent diffusivity values for 95% of the materials but be quite far off for 5% of considered materials.

We discuss some strategies for the effective use of U-MLIPs in their present stage of development in **Sec. 7**. U-MLIPs are also generally slower than targeted MLIPs, as noted in the discussion of MLIP execution speed in **Sec. 6**. That said, there is an enormous advantage to a large, centralized effort around one or a few U-MLIPs. These advantages include the ability to efficiently integrate state-of-the-art improvements, e.g., adding long-range forces, speed optimizations, multi-fidelity learning, fine-tuning, uncertainty quantification, etc. It may be that the aggregation in one place of all the best practices and state-of-the-art approaches helps grow the value of U-MLIPs over targeted MLIPs. Overall, U-MLIPs represent a very exciting advance of MLIPs that will likely have a significant impact on the field of atomistic modeling. The coupled facts that a single potential (i) may soon produce energy, force, and stress values (and perhaps additional properties, such as magnetic moments) with near *ab initio*-level accuracy and order of magnitude more than *ab initio* speed, (ii) can be applied to almost every chemically relevant element in the periodic table, and (iii) can include increasingly complex physics, offers the tantalizing possibility of future U-MLIPs functioning as a truly foundational model for materials modeling, in turn replacing a significant fraction of explicit quantum mechanical calculations with no need for explicit training. Fully realizing the potential of U-MLIPs would allow researchers to quickly and easily explore problems that are practically inaccessible to present physics-based approaches and greatly increase the overall impact of atomic-scale materials modeling.

## 6    Execution (Inference) Speed of MLIPs

Speed for execution of an MLIP is important when one is performing a large number of calculations, which might occur during long MD runs or large-scale searches of configuration and chemical spaces. Key issues to consider for timing are: the processor used for calculations (speed of CPU, GPU, or other hardware), system size, and MLIP type (e.g., complexity, where increasing complexity generally corresponds to greater accuracy and slower execution). It is very difficult to quantitatively assess the speed of MLIPs unless one makes a direct comparison of the same calculations with proper controls for hardware, hyperparameter settings, etc. However, there are some relevant studies available, and qualitative trends can be determined from different performance reports in the literature. We stress that the values given here should be treated very cautiously as qualitative guides for several reasons. Due to the intense development activity, new optimizations are implemented by several groups all the time, and it is likely that in the coming year significant speedups will be obtained by several codes. Careful benchmarking of the latest versions of packages for the particular project should be part of any extensive study where speed is a concern. A common metric for assessing performance that allows for some comparison across different numbers of atoms, processors, and steps from MD or other multi-step simulations is processor-seconds per atom per step, $\left(\frac{N_{proc}}{N_{atom}N_{step}}\right)\tau_{sim}\left[\frac{proc-s}{\frac{atom}{step}}\right]$, where $N_{atom}$ is the number of atoms[88], $N_{step}$ is the number of steps in the simulation (where one step of an MLIP involves evaluation of the total energy and forces on all the atoms for one atomic configuration, e.g., as might occur during one MD time-step), $N_{proc}$ is the number of processors being used), and $\tau_{sim}$ is the wall-clock time required to execute $N_{step}$ steps. The units for each measure are given in brackets. Note that processors could be either individual cores on a multicore CPU or entire GPU processors. Typical nodes on high performance computing resources may contain dozens of cores and up to 6 or more GPUs. As long as each processor has a sufficiently large number of atoms to work with, the performance in proc-s/atom/step will be insensitive to both $N_{atom}$ and $N_{proc}$. The results discussed here will mostly be approximately in this linear scaling regime. An exception to this is the $N_{proc}$=1 special case, where simulations are run on a single core or single GPU. Performance here is usually significantly better than larger scale parallel calculations with $N_{proc}$>>1, where there is additional overhead of MPI network communication. For the $N_{proc}$=1 special case, performance is given in units of simply s/atom/step. All performance results are based on a typical state-of-the-art CPU or GPU from the last few years (relative to 2023). For CPUs, these provide about $10^{11}$ floating point operations per second (FLOPS) and for GPUs (e.g., NVIDIA® V100 Tensor Core GPU) these are about $10^{12}$ FLOPS. Note that in the following discussions we will be giving approximate performance values and thus generally round to the nearest order of magnitude.

First, we consider performance for cases running on a single CPU or GPU processor under close-to-optimal conditions, with a reasonable system size (e.g., 100-1000 atoms) that can fit into memory limits on the CPU/GPU. Good scaling for parallel execution up to very large system sizes has been achieved and will be discussed more below. Timing values for a number of explicit AEF type MLIPs (see **Sec. 4.3**) under different levels of complexity (i.e., basis set size) are shown in **Figure 6**.[1,61] Well-fit MLIPs of the explicit AEF type range from about $10^{-5}$ to $10^{-3}$ s/atom/step (note that this is just proc-s/atom/step for one processor) depending on the number of degrees of freedom used, typically set by the number of terms that are included in the basis function expansions. A typical speed of the faster explicit AEF type MLIPs (e.g., MTP and ACE) is about $10^{-4}$ s/atom/step. As a concrete example of timings, Bernstein evaluated GAP, ACE and MACE potentials for 1024-atom cells of $Cu_xAl_{1-x}$ alloys using MD.[89] He found that ACE timings ranged from about $0.06 \times 10^{-3}$ to $0.18 \times 10^{-3}$ s/atom/step, while GAP was slower at $2.1 \times 10^{-3}$ s/atom/step, both computed for one processor. MACE timings ranged from $0.042 \times 10^{-3}$ to $0.12 \times 10^{-3}$ s/atom/step on a single NVIDIA A100 GPU. For reference, timing is about $10^3$ s/atom/step for standard well-converged DFT in a state-of-the-art code for ~100 atom unit cells of a typical set of elements, $10^{-3}$ s/atom/step for a ReaxFF potential (one of the most complex physics-based traditional potentials), and $10^{-6}$ s/atom/step for the Lennard-Jones and EAM interatomic potentials (some of the fastest physics-based potentials).[90]

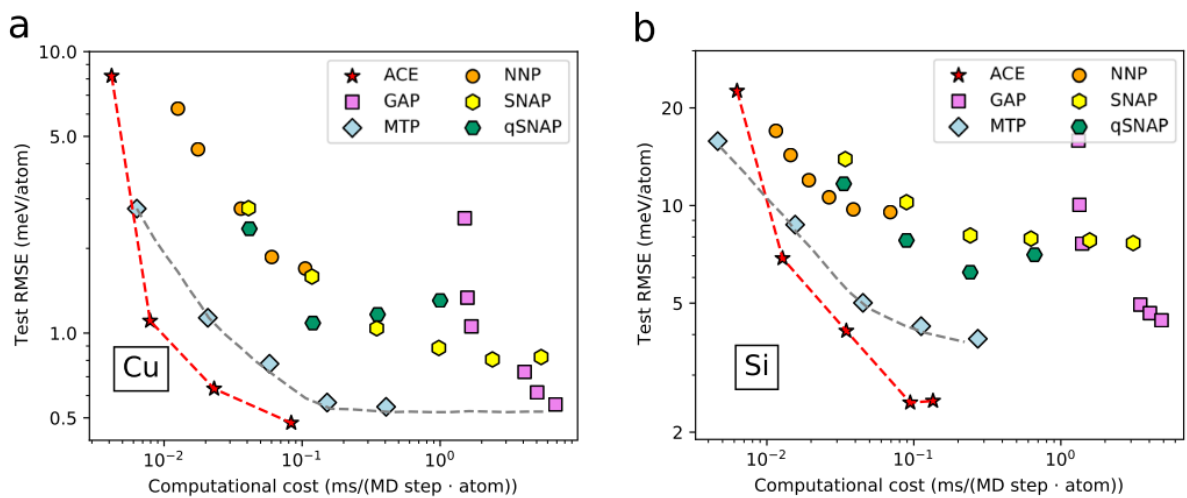

**Figure 6:** Trends in computational cost (speed of the MLIP) for a set of major MLIPs for (a) Cu and (b) Si molecular dynamics calculations (done on one CPU with 108 atoms for 2500 steps). The varying colors correspond to different MLIPs and the points for each color correspond to larger basis function sets, which generally lead to greater accuracy and larger computational cost. The abbreviations are Atomic Cluster Expansion (ACE), Gaussian Approximation Potentials (GAP), Moment Tensor Potentials (MTP), Neural Network Potential (NNP), Spectral Neighbor Analysis Potential (SNAP), and quadratic SNAP (qSNAP). Figure reproduced with permission from Ref. [61] with data originally from Ref. [1].

Next, we consider the timing of deep learning-based MLIPs. Deep learning MLIPs are generally similar to, or somewhat slower, than non-deep learning explicit AEF approaches, although it can be hard to compare as the former are often run on GPUs. Nonetheless, some results exist that give a qualitative sense of the relative speeds of deep learning MLIPs under different conditions. DeepMD typically performs at about $10^{-3}$ proc-s/atom/step on a CPU, and was accelerated ~40 times (so about $10^{-4}$ proc-s/atom/step) on a GPU in a direct comparison.[91] M3GNet,[41] which models a very large number of elements (89) and is what we refer to as a U-MLIP (see **Sec. 5**), takes about $10^{-3}$ proc-s/atom/step on a single CPU to perform a structural relaxation of $K_{57}Se_{34}$, an example chosen for its large energy change during relaxation. Recent testing of the PFP U-MLIP from Matlantis gave about $10^{-3}$ s/atom/step on a GPU for 100-1000 atom unit cells running MD in LAMMPS.[92] These values were about 5 times slower than well-converged MTP and ACE fits on identical systems run on a single CPU, and about 50-100 times slower than the same runs on a large set of CPUs.

An interesting developing area to increase MLIP speed is the ultra-fast approach,[42] which uses computationally cheap spline functions to describe the atomic environments and linear regressions for energy/force predictions. The potentials resulting from the ultra-fast approach are extremely fast compared to existing MLIPs at the price of limited flexibility and possibly greater errors for complex systems. For example, such potentials are about $10^3$ times faster than typical explicit AEF MLIPs, with similar prediction accuracy to SNAP, GAP, and MTP on some test cases, putting them at about $10^{-6}$ s/atom/step and comparable to the fastest simple PBPs.

Efficient architectures and scaling up the number of CPUs and GPUs used to evaluate MLIPs can lead to large speedups, which is particularly useful for the somewhat slower deep learning methods. Note that these timing values are somewhat faster than above, likely because the inclusion of more atoms is allowing for more efficient use of the processors. DeepMD achieved about $10^{-5}$ proc-s/atom/step with about 127 million Cu atoms, and SNAP achieved $10^{-6}$ proc-s/atom/step on about 20 billion C atoms, both running on 27,900 GPUs (4650 nodes on the Summit machine).[93] A deep learning MLIP particularly optimized for scaling and performance is Allegro,[13] which uses a local equivariant neural network and ACE-like atomic features, and while it can be executed on CPUs, it is best run on GPUs. Allegro models of water achieved about $10^{-5}$ proc-s/atom/step with 4, 64, 1024 GPUs and $10^5$, $10^6$, $10^7$ atoms, respectively.[94] Note that the choice of hyperparameters (i.e., complexity) can change this approximate timing by an order of magnitude and that this is for an optimally tuned MLIP. To increase execution speed, typically a more complex model is first used to verify the fidelity of the training data and learning process, before being reduced in size, while still reproducing a target property of interest with sufficient accuracy.

As another example of an MLIP particularly optimized for scale and speed, the GPU implementation of the FLARE potential,[25] based on C++ with a Python wrapper, was used to model heterogeneous catalysis of $H_2$/Pt(111) for 0.5 trillion atoms on 27336 GPUs nodes, achieving $10^{-6}$ proc-s/atom/step.[95] However, the speed of FLARE on CPUs reduces significantly compared to GPUs, down to roughly $10^{-3}$ proc-s/atom/step. It is useful to note that the performance for SNAP, FLARE, and Allegro all begin to deviate significantly from linear scaling of inverse time with processors (constant $B$ values) by around $10^5$ atoms/GPU for the Summit hardware used in these tests (NVIDIA V100-16GB GPUs). While these timings are very impressive, it is important to realize that PBPs can also take advantage of parallelization and GPUs. For example, a GPU-accelerated classical force fields model based on the Martini potential recently achieved 6 microseconds/day on 136,000 particles ($B=10^{-9}$ proc-s/atom/step) using six V100 GPUs.[96]

In summary, from the above-discussed timings we can learn at least two important lessons. The first lesson is that, similar to PBPs, scaling up to even hundreds of billions of atoms is possible for some MLIPs. These calculations generally require multiple GPUs, which can be a challenge to access, but options are becoming increasingly available (see discussion of infrastructure for MLIPs in **Sec. 8**). The second lesson is that the general trend of speed we noted on one CPU, which is that simple PBPs are fastest, followed by explicit AEF MLIPs, then finally implicit AEF deep learning MLIPs, largely still holds with larger-scale calculations. That said, we stress that the details of the MLIP fit and optimization can matter a lot for large-scale calculations, so one should choose an optimal approach carefully if pursuing such studies.

# 7   MLIP Choices – What Should I Use When?

When choosing MLIPs, many factors can be considered. We list a few of these factors in this section and provide some guidance on how to think about each of them. We start from basic aspects of hardware, accuracy, and speed and then progress to the details of pursuing a specific MLIP.

## 7.1   Hardware Resources

Hardware resources could be an initial deciding factor in choosing MLIPs both when fitting a new potential or using a pre-trained potential. Generally, explicit AEF MLIPs such as MTP,

ACE, SNAP, and GAP have fewer parameters and functions than NN-based MLIPs and run well on CPUs. On the other hand, NN- and GNN-based MLIPs mostly rely on GPUs. Some potentials, like MTP, can presently only be run on CPUs, while ACE is faster when fit using GPUs but can be used for MD simulations on both CPUs and GPUs. NN-based MLIPs are primarily created to be fit and used with GPUs, although they can be run on CPUs, with typically a 10-100× slowdown on CPU vs. GPU calculations (see discussion of MLIP timings in **Sec. 6**). These trends generally suggest that if you only have access to CPUs, then explicit AEF MLIPs are likely best as they will be certain to run and will typically run with reasonable speed. If you have access to GPUs, then both explicit AEF and NN-based MLIPs are potentially good choices. Given the growing importance of U-MLIPs and the use of GPUs in training and executing many MLIPs, it is likely advisable to have access to at least one high-performing GPU if you are planning extensive use of MLIPs. In addition to the discussion above, in Matlantis, which is commercially deployed as SaaS, PFP is provided via an API, allowing users to execute inference without considering the environment setting and optimization of computing devices. In practice, the inference is executed in backend GPUs or specific deep learning accelerators named MN-Core series. [97]

## 7.2 Speed Requirements

The overall simulation time depends on the size of the system, the number of execution steps in the simulation, available hardware resources, and the computational cost of the MLIP. Assuming the first three factors are fixed by the project and infrastructure needs of the user, the MLIP framework determines the overall simulation time. As previously discussed in **Sec. 6**, explicit AEF MLIPs are about 10-100× faster than implicit AEF deep learning MLIPs. If the computational cost is not a limitation, deep learning MLIPs typically provide the highest accuracy and may be adopted. Otherwise, the user could opt for any of the explicit AEF MLIPs that provide the desired accuracy.

## 7.3 Accuracy Requirements

While the promise of MLIPs is to achieve any desired property accuracy with respect to *ab initio* methods, in practice there is an accuracy limit of MLIPs to keep the computational cost of the MLIP reasonable given the available resources. One limiter of MLIP accuracy stemmed from the insufficient description of the atomic environment in the earlier MLIPs such as Behler-Parrinello NNs, GAP, and SNAP: the fact that three-body (and even four-body) descriptors of local environments are incomplete was shown explicitly.[98] More recent MTP and ACE formalisms introduced new methods to give a systematically improvable description of the atomic environment using high body order basis functions and used linear regression to learn the PES, enabling an increase in the accuracy of MLIPs while keeping the computational cost tractable. In recent years, it has been shown that equivariant GNNs can achieve very high accuracies with a practical computational cost, where NequIP, Allegro, TeaNet, SO3krates[99] and MACE are examples of such approaches. Thus, the authors recommend that if GPUs are available, training with equivariant models such as NequIP, Allegro, TeaNet, SO3krates or MACE is likely to yield the highest accuracy. Furthermore, it has been found that higher accuracy may be obtained by fine-tuning a pre-trained potential as opposed to to training a new MLIP from scratch, even for tasks that were out-of-distribution compared to the training data.[81] In a few personal experiences by the authors, we have found that for real systems, with abundant training data available (meaning we could keep running more DFT as needed until we see little improvement in the potential), the AEF methods like ACE tend to have root mean squared error (RMSE) on energies and forces that are 2-3 times those of GNN methods like MACE. If there are only CPUs at hand, the authors suggest MTP or ACE for the moment, until GNN packages are implemented for CPUs efficiently in the future. Implementations of these various methods are likely to improve and diversify utilizing popular hardware, and therefore we expect the field to change rapidly.

### 7.4  Using A Pre-trained Potential

Depending on the type of study, one may decide to use a pre-trained potential or to fit a potential from scratch. It will likely save a lot of time if one can start from a pre-trained potential, so this is a logical first step to explore. Pre-trained potentials may be found in online repositories or by searching through scientific articles. For example, pre-trained targeted MLIPs for specific systems (e.g., GAP potential for Cu) can be found on the NIST Interatomic Potentials Repository and the Open Knowledgebase of Interatomic Models (OpenKIM).[100–102] When deciding to use a pretrained potential, one must make sure that the potential is suitable for the study. Given the recent availability of U-MLIPs and their ease of use across many systems, they represent an appealing option, and importing pre-trained versions of U-MLIPs from their respective repositories is straightforward.[103–105] However, although U-MLIPs generally have low energy and force errors compared to their *ab initio* training data, their ability to predict accurate materials properties is not ensured by these low errors (see **Sec. 5**) and they have not been thoroughly validated for accurate prediction of materials properties across most systems. It is therefore quite possible that despite some impressive successes (see **Sec. 5**) that many properties, from vacancy formation energies to melting temperatures, may be incorrectly predicted. Furthermore, U-MLIPs can be slower than other MLIP or PBP approaches (see **Sec. 6**), so speed requirements should be considered. However, given the rapid rise of such U-MLIPs in just the past couple of years, it is likely that increased property prediction benchmarking will be available, and iterative improvements to the U-MLIPs, e.g., through fine-tuning, will further aid in improving their accuracy and generalizability. For the time being, there are a few simple strategies one can use to apply U-MLIPs most effectively, which we summarize here:

1. Validate the U-MLIP predicted energies and forces for your system of interest. One way to ensure the accuracy of an untested MLIP for a specific system and purpose is to run some relevant *ab initio* simulations for your problem and compare the *ab initio* and U-MLIP energies and forces. One should be careful to choose *ab initio* settings such as functional, energy cutoff, k-point density, etc., consistent with the training data used for the MLIP (e.g., choosing the right pseudopotentials and Hubbard U values for GGA+U calculations). Good agreement is strong support that the U-MLIP is applicable to your system. Such a benchmark can be done with just a handful of static *ab initio* calculations on small unit cells and therefore can be quite fast. The use of benchmarks that directly relate to the property of interest, e.g., an activated state for a chemical reaction or few points on a gamma surface for stacking fault energies, are likely best.
2. Validate the U-MLIP property predictions for your system. In many cases, the benchmarking described above can be easily extended to include comparing *ab initio* and U-MLIP calculation of specific properties of interest, e.g., a set of phonon dispersion curves, diffusion coefficients or defect formation energies, particularly for small systems or simplified cases. Good agreement on target properties provides even greater confidence in the U-MLIP than just similar energies and forces on select structures. For example, one might calculate diffusion coefficients in a small unit cell with *ab initio* and the U-MLIP and, if similar results are achieved, apply the U-MLIP to much larger systems or different compositions.

3. Apply U-MLIPs to problems that can easily detect failures or are not overly sensitive to failures. Many applications might not suffer too much from intermittent failures of the U-MLIP. For example, using a U-MLIP to pre-relax other *ab initio* calculations is a very robust application tolerant to U-MLIP failures since the final calculated result does not directly depend on the accuracy of the U-MLIP. In addition, failures in the pre-relaxing can be easily identified and corrected by checking against the corresponding *ab initio* relaxation. As noted above, in developing the M3GNet U-MLIP, Chen and Ong comment that pre-relaxing hypothetical structures with their U-MLIP before performing *ab initio* calculations resulted in approximately 3× time savings compared to running *ab initio* on un-relaxed structures.[41] Another example might be using U-MLIPs for an initial screening of a large set of candidate materials for a specific property, where a highly accurate calculation may not be necessary in the initial steps. Failures of the U-MLIP might lead to false positives (keeping unpromising materials) or false negatives (removing promising materials) but later screening with full *ab initio* calculations can catch the false positives, and, typically, screening is often more focused on getting a few successes than ensuring no false negatives. A final example is generating physically relevant atomic configurations (which need to be calculated with *ab initio* methods later) for training a more specific potentials, a way in which U-MLIPs might help accelerate the development of more targeted MLIPs.

Despite the exciting potential of U-MLIPs, the high levels of uncertainty in their applicability means that many practitioners presently still either fit their own potential or use a pre-trained potential that is specifically fit for the material under investigation. As a new trend different from this, some early adopter researchers have begun to perform calculations without finetuning. For example, Matlantis provides pretrained U-MLIP (PFP), with the aim of allowing users to do practical simulations without having to perform finetuning. In all cases, it still matters that the training data used for the potential is consistent with the type of study being considered, both in terms of atomic structures, chemical states, and relevant physics. For example, (i) an MLIP that is trained on pristine crystalline phases and crystals with stacking faults and vacancies may not be appropriate to conduct a study on the amorphous phases of the same material, (ii) an MLIP trained on low valence transition metal states might not represent high valence states of these same metals well, or (iii) a potential trained on *ab initio* methods like the DFT-PBE functional may not be suitable for layered materials or molten salts, where vdW contributions are significant (although in this case the potential might be corrected by adding empirical vdW corrections). In general, for all MLIPs, one should validate the energy, force, and property predictions as much as possible for a specific use case unless it very closely matches previously published or well-validated work. Many considerations related to the issues above are likely relevant for choosing an optimal pre-trained potential but, since the potential has already been developed, it is likely that the original authors have already taken these items into consideration (e.g., choosing the right potential for the hardware they ran on, etc.). Thus, one can take guidance from the earlier work about optimal use. That said, it might still be useful to have a sense of how different potentials behave related to the above issues, and in the next section we summarize the key concerns in the context of fitting a new potential.

## 7.5 Fitting A New Potential: General workflow

### 7.5.1 Basic ideas

If no pre-trained potential is available, one will need to choose an MLIP framework and fit a potential from scratch. In this section, we provide some strategies and guidance from hands-on experience to help new users approach choosing an MLIP to fit. In addition to the above considerations when using a pre-trained potential, a few new factors become relevant when fitting your own potentials, which we discuss here.

The basic idea behind fitting MLIPs is the same as in almost all regression ML problems. One defines a loss function and adjusts the parameters of the ML model, typically using some kind of matrix inversion or backpropagation, until the loss function is minimized. For MLIPs, the loss function is usually a weighted sum of RMSEs on a few targets, which are usually forces on atoms and total energy, but can also include other properties such as stress tensor, virial, polarizability, etc. Typically, the most important terms are the RMSE in forces and energy and these are standard to report. It is important to realize that, although MLIP fitting is similar to other ML models, it is helpful to use domain knowledge (from physics, chemistry, and materials science) to perform successful training and assessment, which we call *science-informed fitting*. Science-informed fitting is very helpful, at least at present, because the MLIP fit will almost certainly not be perfect for all possible configurations of atoms, so the user is suggested to apply their domain knowledge to develop a model that is adequate for their needs.

At present, there is no agreed-upon standard or widely accepted optimal workflow for fitting an MLIP. However, multiple authors have provided very helpful articles that cover the major considerations and provide excellent practical guidance.[106–109] Here, we describe the typical general workflow, and then go into some of the detailed questions and choices associated with its implementation. In addition, a standard set of procedures and software for generating or acquiring training data, fitting, comparison, and deployment of MLIPs is provided in **Sec. 8**. The general approach is to generate an initial set of *ab initio* data $\{s_1\}$, consisting of atomic configurations related to your problem of interest (e.g., liquid configurations for studying a melt, different vibrational modes for studying phonons, or multiple distortions for studying molecular systems). Then, fit an initial potential to ~80% of $\{s_1\}$ and test on the left-out ~20% of $\{s_1\}$ to assess accuracy on energy and forces (this approach and the details below can be readily extended to other targets if they are used). In the case of training GNNs, it is common practice to train on 80% of the data, reserving 10% for validation (to guide the GNN training process) and 10% for testing. If the fit quality is not adequate (e.g., the force and/or energy RMSE is too high), one develops additional data, adds it to $\{s_1\}$ to form a new data set we call $\{s_2\}$, and then performs a similar assessment. If $\{s_1\}$ is sufficiently large, then no iterations may be needed. If the system is complex and/or relatively small data sets are being added at each step, then this might take many iterations. Atomic configurations for different $\{s_i\}$ are generally determined based on user intuitions about important configurations for the application of interest (e.g., known stable compounds in the material), independent samples from MD trajectories, guidance from active learning (discussed below), or some combination of all of these, depending on the application. The required amount of data to obtain a desirable fit can vary, but for typical systems with 3-4 species, the number of total energies $N_E$ and the number of forces $N_F$ used in training are approximately $N_E \sim 10^3$ and $N_F \sim 10^5$. This estimate is very approximate, and model type and architecture (e.g., MTP vs. ACE, equivariant vs. invariant features, etc.) can also affect the results. In particular, for deep learning methods, the error vs. amount of training data (the learning curve) is expected to follow a power law, but the power law exponent can depend on significantly on the details of the MLIP.[110]

### 7.5.2 Determination of test data

The first potential fitting issue we address is strategies for determining useful test data sets for validating the MLIP fit. Random cross-validation (CV) or k-fold CV are both reasonable if the data is not highly correlated. However, if the data has many similar conditions, e.g., as occurs for data generated from AIMD trajectories or small perturbations to existing structures, then these random CV approaches will yield overly optimistic predictions. The predictions will be overly optimistic due to the "twin" problem, where extremely similar data is present in both the train and test sets, and the model predictions are thus indicative of data that looks just like the training data. In the case of highly correlated or otherwise similar data, one can assess the potential more robustly by comparing *ab initio* and MLIP predictions from new conditions, e.g., MD at a new temperature or MD from a much later time than that used during training. An even better way to assess the MLIP in such cases is to apply the MLIP in expected or near-to-expected use cases and check errors on select configurations from those conditions. For example, assume you are trying to predict the diffusion of Li in a solid-state electrolyte at low or even room temperature. The bulk of the training data might be AIMD trajectories at higher temperatures so that many Li hops occur. An example of good test data would be to simulate low-temperature hopping with the MLIP, extract configurations where the hopping occurs, run these with *ab initio* methods, and compare the *ab initio* and MLIP energies and forces for those configurations as a test. Obviously, when possible, testing the ability of the potential to predict the properties of interest is essential. Continuing the example above, one should be sure that the *ab initio* and MLIP-predicted Li diffusion match in the higher temperature conditions where the *ab initio* simulations are reliable and can be well-converged. However, extensive property testing is generally difficult as it can be challenging to have a robust ground truth, proper simulations often take a long time for the ground truth even using the MLIP, and there are generally relatively few property values for comparison (e.g., one might have only 5-10 densities or diffusion coefficients as compared to many thousands of forces). This disparity makes it desirable to know as much as possible that a potential will be robust before starting significant property exploration. This robustness is generally assessed through energy and force errors and brings us to the second issue.

### 7.5.3 Required energy and force accuracy

The second potential fitting issue we address is what accuracy of energies and forces is needed in the test data to ensure a useful MLIP, by which we mean an MLIP that can be used for a wide range of simulations and yields accurate predictions for properties of interest. At present, there is no exact answer to this question. The accuracy that can be achieved will depend on the conditions being explored and the range of elemental species and structures considered, as well as the type of MLIP. For example, a simple liquid phase of just one element may yield much smaller errors, both relative and absolute, than modeling oxidation of a complex surface at high temperatures. However, there are still challenges in learning even single-element systems. For example, Owen, et al.[111] found that early transition metals have higher relative errors than late platinum- and coinage-group elements. This apparent difficulty in learning is attributed to the sharp *d*-electron density of states above and below the Fermi level, resulting in complex physics which makes the PES difficult to learn. The relative energy and force errors in their study of transition metals ranged over about a factor of 10. That said, typical values for energy and force errors are in the ranges of 1-10 meV/atom and 20-40 meV/Å, respectively, for a very good fit, although force errors of up to around 100-200 meV/Å have been reported in nominally successful MLIPs.[1,111] Very accurately trained MLIPs can achieve errors for energies, forces, and stress tensor components on the order of 1 meV/atom, 10 meV/Å, and 0.1 GPa, respectively, although in practice one may find somewhat higher (e.g., 2×) energy, force and stress errors which are highly system and potential dependent.

It is reasonable to assume that for a relevant and diverse set of training data, a lower RMSE on energies and forces will generally translate into more accurate property prediction. However, depending on the application, low energy, force, and stress errors may not be sufficient criteria for ensuring accurate property predictions.[112,113] In addition, an MLIP trained on a large number of chemically diverse systems may exhibit energy and force RMSEs that vary widely by element or chemistry type (e.g., defects in oxides vs. elemental metals), system state (e.g., solid vs. liquid), and simulation conditions. Obviously, the MLIP is at best as accurate as the *ab initio* method used to train it, so for the discussion here we will assume that the *ab initio* method yields accurate results. In the case of negligible RMSE on all atoms in all situations, it is expected that the MLIP is essentially equivalent to the *ab initio* method used to train it and will ideally yield robust property prediction. However, this ideal scenario is difficult to reach in practice, due to poor predictions on outliers. RMSE values are averages over many configurations, so even MLIPs with low RMSE can have outliers that have significant errors. If these outliers are important for a given property, then the prediction may not be accurate. Again, referring to the example above, an MLIP trained on a large body of *ab initio* MD simulation data of a Li conducting compound may show very low RMSEs on energies and forces on all the different atom types, but still not accurately capture the activated state energy of Li during a hop (i.e., this activated state is an outlier) and therefore yield inaccurate diffusion coefficients. The result of a low RMSE but the inability of the model to capture some piece of physics is analogous to situations that often arise when developing standard ML regression models, where the model is generally reliable for interpolation tasks (test data similar to training data) but unreliable for other tasks, even when not formally extrapolating.[86]

As a concrete example of MLIP extrapolation issues encountered during a study, Zhai et al.[114] demonstrated that a widely-used deep neural network potential, i.e., DeepMD, can reliably reproduce the properties of liquid bulk water but provides a less accurate description of the vapor–liquid equilibrium properties. This problem can be compounded by two potential issues: (1) The ML architecture cannot capture the essential symmetries and physics, e.g., many-body interactions; (2) The training data is not at all evenly distributed in structural or chemical space, a common issue when data is sampled from MD or biased toward widely studied compositions, leading to data imbalance issues when training robust MLIPs. As discussed above in the hypothetical case of studying Li conductors, the simplest way to avoid such issues is to be sure that the training data samples as much of the relevant configuration space as possible. If one is concerned about predicting diffusion, then use training data with many activated states for hops, and if one is concerned about predicting bulk moduli, then use training data from a range of different stresses. Another way to improve predictability is to change the evaluation metrics to include force predictions on important outliers. This technique was suggested by Liu et al. when they observed that large discrepancies can still be observed in migration barriers even when defects are included in the training.[115] Considering relevant rare-event-based metrics (e.g., accuracy for diffusion hops, defects, atomic vibrations) for MLIPs is important, since it is for these configurations where force errors can potentially be large.

An additional complexity in ensuring a robust potential is that small RMSE is not a guarantor of stable simulations.[113] By *stable simulations*, we mean particularly long-time (e.g., tens of nanoseconds) MD simulations.[116] There is a tendency for MLIPs to become unstable during MD simulations and crash. Depending on your needs, this can make the potential useless. We hypothesize that crashing of the potential typically occurs due to the system exploring regions of configuration space where forces are not accurate and change in ways that are too fast for the MD time step to manage. This leads to errors that accumulate and eventually cause numerical instability. In other words, the numerical integration of the equations of motion being performed by the MD becomes unstable because the energies and forces are, at least during some parts of the simulation, not changing slowly on the time scale of the MD time step. Such an event is not unlikely if the potential becomes unphysical, since the MD time step, generally taken to be 1-2 fs, is tuned to be effective for a physically realistic system. This problem can be reduced by starting with progressively more varied training data. It can also be remedied by running *ab initio* calculations on configurations from the MLIP simulation just before the observed instability to obtain new training data, which can stabilize the model after retraining. It is also possible to flag configurations that appear during the use of the MLIP that are in some way outside the domain of the training data and running *ab initio* calculations to add these cases to the training data. The domain of the training data is typically determined using some measure of difference from the training data, e.g., active learning with D-optimality (discussed more below).[117,118] These domain-based approaches are quite effective in establishing a stable potential for MD and are widely used. Such approaches may require multiple iterations, and it is not clear *a priori* how many will be needed to achieve a stable simulation, although typically no more than 5 iterations are needed. The above discussions offer many qualitative guides for training and test data, but do not provide any concrete approach to assembling a training database, which brings us to our discussion of this third important issue.

### 7.5.4 Determination of training data, use of active learning

The third potential fitting issue we address is how one should choose training data. Again, this does not have a unique settled answer, but there are useful guides. The simplest approach is to use domain-specific intuition to develop a training database that is diverse, relevant, and large. This is easier than it might sound, and given the speed of modern *ab initio* methods, often quite practical. The advantage of this "intuitive structures" approach is that it is relatively easy to implement, makes good use of materials knowledge, and tends to yield a good MLIP in a practical amount of time. However, the approach is almost certainly not optimal in terms of getting the best potential for the least training data, it is not readily automated, and it may not scale well to MLIPs that are targeting many elements and many kinds of physics all at once. A different second approach that seeks to solve these issues is active learning, which is described next.

For the most efficient training data generation, users have a few options, and *active learning* is commonly useful. Note, by active learning we mean an iterative approach that uses the results of a collection of fits to suggest the best new training data to add for the next fit to optimize some condition, e.g., creating an MLIP with the lowest RMSE on some property. A general overview of the use of active learning to train MLIPs is provided in **Figure 7**. To apply active learning, models need access to uncertainty estimates during configurational sampling. When uncertainties are high, a ground truth calculation (i.e., *ab initio* calculation) is automatically invoked. The most common active learning approaches for MLIPs are D-optimality,[118] Gaussian process regression,[25] querying by a committee of GNNs,[119,120] Bayesian inference force fields,[121–123] and uncertainty-driven MD simulations with bias potentials.[124,125] The active learning process typically makes use of the featurization of the atomic environment to automatically guide the search for unseen and uncorrelated atomic configurations to improve model predictability. Implementing active learning therefore requires access to the featurization used in the MLIP, and is most easily applied when built into the MLIP package. For instance, MLIPs such as MTP, ACE, and FLARE have built-in active learning functionality in their fitting routines. It is possible to use a featurization separate from the MLIP to implement one's own active learning framework. Packages such as Dscribe[126] and matminer[127] can be used to featurize the atomic configurations for developing one's own active learning or other data generation approach.

Here we give a few examples of fitting approaches used in recent studies. Attarian et al.[128] explored the intuitive structures vs. active learning based on D-optimality in a study of properties of eutectic composition FLiBe salts with an MTP potential. They found that either way of training data generation resulted in a robust potential, though the active learning approach was more efficient as it produced about the same prediction error with less than half as many training structures (600 vs. 1400 structures). Work from Vandermause et al. also compared the use of active learning vs. random sampling, and they found that active learning resulted in more efficient MLIP training (i.e., lower RMSE per training data added) and an overall lower RMSE compared to random sampling.[25]

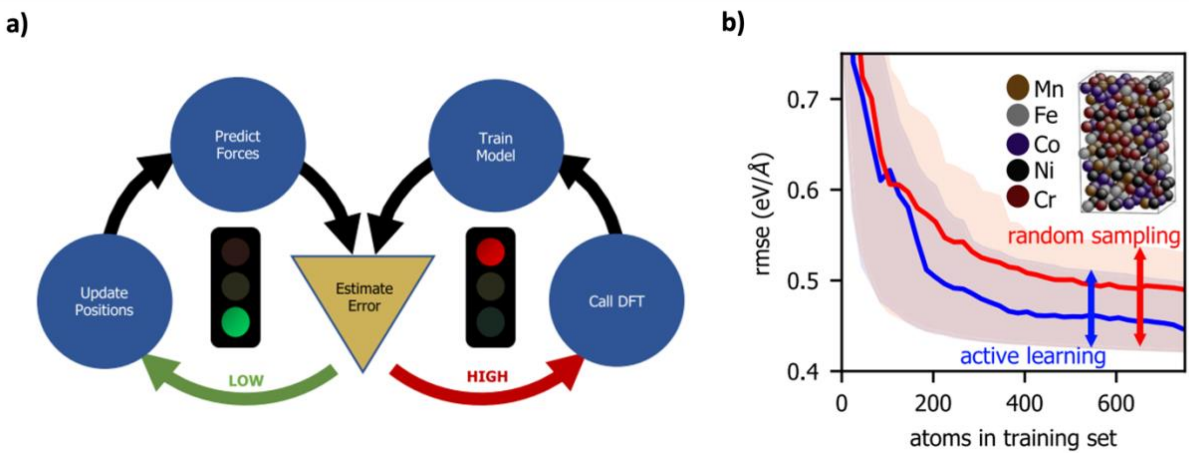

**Figure 7.** (a) Overview of the use of active learning in constructing reliable MLIPs. (b) Learning curve showing improved efficiency of active learning approach vs. random sampling for developing MLIP of 5-component high entropy alloy. Adapted with permission from Ref. [25].

There are additional approaches for the efficient generation of training data that do not leverage active learning. For example, in a recent study using bias potentials, Kulichenko et al. merged the ideas of querying by committee and metadynamics to model the phase space of proton transfer in glycine.[124] The use of a bias potential, instead of high-temperature MD simulations, generates low and high-energy configurations, thus avoiding sampling unnecessary structural distortions. When the main purpose of active learning is to add weakly correlated or uncorrelated configurations to the training data, bias potentials may be a direct and efficient approach. As a second example, in their study of Mo, Chen et al. outlined the selection of training structures using principal component analysis, and the selection of hyperparameters using a differential evolution algorithm.[129] Their procedure, using a SNAP MLIP, achieved close to DFT accuracy for elastic constants, melting point, and surface and grain boundary energies. As a third example, Vandermause et al. built a FLARE potential for vacancy and adatom diffusion in Al, where the training data was obtained on-the-fly, where select DFT calculations were performed if the GPR error bar became too large. In their MD runs, they found that the majority of the DFT calculation calls occurred near the beginning of the run, with no DFT queries occurring after 400 ps of MD time.[25]

A valuable tool for developing training data can be to use a classical PBP or a U-MLIP to generate a large initial set of atomic configurations, which are then sampled intelligently to obtain DFT runs for training data. This sampling can be done with active learning, as described above. It can also be done with other approaches. For example, a clustering algorithm can be used to separate different groups of configurations based on some similar features and later a collection of configurations from each cluster is chosen to be calculated with *ab initio* methods and used as the training set. Users can take advantage of packages such as Dscribe[126] to featurize the atomic configurations and ML packages such as scikit-learn to do the clustering. More recently, enhanced sampling techniques have been utilized to accelerate the sampling of rare events and integrate that sampling with active learning procedures for the generation of training datasets for MLIPs that can describe rare events.[124,125,130]

## 7.6 Fitting A New Potential: More specific considerations

### 7.6.1 Chemical Complexity

As discussed in **Sec. 4.3**, a drawback of most explicit AEF MLIPs as they are currently formulated is that they scale poorly with the number of species. This scaling results in a higher computational cost for systems with a higher number of species both when training and executing simulations with the potential. For example, FLARE is generally extremely fast in its execution time (see **Sec. 6**), but can scale poorly with training set size and chemical complexity. The general rule is that an update (i.e., retraining of model parameters) of the FLARE sparse Gaussian process can become prohibitively expensive when there are around $N_{env}$ = 1,000,000 environments in its training set, where $N_{env} = O[$ (# training *ab initio* frames) × (# atoms/frame) × (# species)$^2$] , where a frame is one set of *ab initio* calculated forces and energies (note that 1M environments is an upper bound, where in practice FLARE users may experience slow timing and large memory requirements when using >600k environments).[123] The quadratic scaling with the number of species is particularly limiting for chemically complex systems since explicit AEF MLIPs can handle less training data, but typically need sufficiently varied training data to explore the many chemical configurations. Considering the above limit for FLARE as a concrete example, offline training (where *ab initio* calculation frames have already been calculated and are available for fitting) for a system with a single species is possible for about 4000 training frames (250 atoms per frame) while for 5 species the scaling limits the user to about 150 frames, which is likely too few to fit an accurate potential. Similar issues exist for MTP, ACE, and other explicit AEF MLIPs, and a brief review of the literature shows that almost all fits with these explicit AEF MLIPs are to 5 or fewer chemical species. A recent attempt by the authors to fit an ACE potential to a 12 species system of chloride salts with 3500 training data using Nvidia Tesla v100-32 GB GPU failed at the very beginning and the code did not even start the training. A more in-depth discussion of scaling issues is given in **Sec. 4.3**. [29]

### 7.6.2 Training Requirements

The difficulty of training an MLIP is a key factor in choosing one that is right for your project. Key things to consider include both the amount (and potentially variety) of training data and the training time. All other things being equal (e.g., for the same chemical system and desired accuracy), the training data requirements for different potentials can be quite different. For example, molten salt FLiBe potentials were recently trained with DeepMD and MTP.[128,131] Both approaches produced excellent potentials, but the MTP fitting was achieved with less than 1% of the amount of the DeepMD data (although it should be noted that this was not a head-to-head comparison by the same authors under identical conditions so should be taken as only a qualitative guide for the training data differences from these potentials). Early deep learning MLIPs[131,132] required much more training data than explicit AEF MLIPs, but this no longer seems to be true for the newer equivariant deep learning MLIPs, which are much more data efficient. For example, studies with NequIP report a 1000× improvement vs. DeepMD with respect to data requirements.[133] However, even if a deep learning and explicit AEF MLIP require the same amount of training data, the complexity of the former will typically cause it to train more slowly.

### 7.6.3 Ease of Fitting (Tools and Hyperparameters)

Ease of the fitting process is also a key factor in considering which potential to use. First, it is important to have good fitting tools associated with the potential that allow for easy fitting, ideally with active learning. Most popular potentials now provide such tools, and more are being developed rapidly, so we will not say more about this requirement and just assume it is satisfied for any potentials one might consider. More fundamentally, potentials with fewer hyperparameters are significantly easier to use. This difference can be large, ranging from just one hyperparameter in MTP, which makes hyperparameter optimization trivial and fast, to many for Allegro, which can require significant experience and skill to optimize to achieve state-of-the-art results. In this regard, the authors have found that MTP is one of the easiest MLIPs to fit as it only has one hyperparameter, which is called the "complexity level" parameter of MTP, and the user can start from lower levels and increase the complexity level step-by-step to achieve the desired accuracy. It should be noted that as of this writing (early 2024), MTP does not support GPU training, so with large training sets many CPU cores are required. However, this hardware limitation may be removed at any time with an update to the MTP code. Compared to MTP, the ACE potential provides much more flexibility in terms of fitting parameters for interaction between each pair, triplets, etc., of species. This flexibility creates a lot of hyperparameters, which correspond to the bond order of many-body interactions, the number of radial basis functions, and the angular resolution of the description. However, because ACE featurization allows for good physical intuition, after a few training sessions with different hyperparameters, the user gets an understanding of how to balance the hyperparameters to achieve the desired accuracy while keeping computational cost low. While this modest complexity from hyperparameter optimization may seem unimportant, it can easily increase the overall time to fit a potential by a few multiples. This is because the *ab initio* simulations and fitting efforts are largely automated, but the hyperparameter optimization is still often done somewhat sequentially and by hand. This challenge may reduce quickly as standardized hyperparameter choices emerge or more automated optimization methods become available.

### 7.7 Summary of Considerations for Choosing a Potential

There are many options for possible MLIPs for fitting, including those already mentioned in this paper as well as many others, and, as with many aspects in this emerging field, there is no standard consensus on the best MLIPs. However, we can provide some guidance to help users navigate the options. If one needs a fast potential (e.g., simulations for tens of nanoseconds and longer) and/or one does not have access to GPUs, then explicit AEF MLIPs are likely a good starting point, where we suggest starting with MTP or ACE due to their ease of fitting and high accuracy, respectively. Conversely, if one does not need a lot of speed (e.g., exploring a few thousand structural energies) and/or one has access to GPUs, then deep learning potentials are a practical option, although not necessarily required or even the best option. The requirements of accuracy, noted in **Sec. 7.3**, suggest using the more complete potentials (e.g., MTP, ACE) vs. the older forms (e.g., SNAP), due to greater potential accuracy with no obvious downsides. In particular, the work of Zuo, et al. performed a very useful comparison of different MLIPs in 2020 and found that MTP was both highly accurate and very fast to execute, performing generally somewhat better than GAP, SNAP, and Behler-Parrinello NN potentials.[1] This suggests that MTP is a good potential to start with in the absence of more information. As discussed in **Sec. 4.3**, recent developments in MLIP formalism have shown that essentially all of the basis functions that underlie different explicit AEF methods (e.g., ACSF, SOAP, HBFs, MTFs) are special cases of the ACE formalism.[14] This suggests ACE is a method of choice, but its flexibility comes with more hyperparameter choices, which can make it more complex for the user to navigate.

Using a pre-trained MLIP avoids training time, which is typically days to months, and is therefore worth pursuing (**Sec. 7.4**). U-MLIPs can be a great starting point, but need to be carefully vetted, and at this stage are likely best used in cases where some post-calculation checking is built into the project workflow. Reusing targeted MLIPs can be a great solution, but it is recommended to validate at least some aspects of the MLIP since it is likely being used in some ways different from those in the original studies and assessments. Finally, if you are fitting your own MLIPs, then deep learning potentials are generally needed for more than ~5 elements, although recent methodological developments are potentially removing this constraint (**Sec. 7.6.1**). However, deep learning MLIPs can require more data and take more time to train (**Sec. 7.6.2**), and may require more human time for hyperparameter optimization (**Sec. 7.6.3**).

## 8  MLIP Infrastructure

In recent years, a plethora of software packages have emerged within the dynamic landscape of MLIPs, catering to both standard-scale and large-scale simulations. These packages aim to streamline the process of training, fitting, and deploying MLIPs for running MD for diverse applications in chemistry and materials science. For standard-scale simulations, ideal features for packages should include ease of use, adaptability, intuitive interfaces, and flexibility in handling various data types and model architectures. On the other hand, large-scale simulations demand efficient parallelization, scalability, robustness, and high-performance computing integrations. In the following section, we explore some of the most prominent and user-friendly packages in both categories, detailing their features, strengths, and ideal use cases.

In the effort to promote adoption and advance the accessibility of MLIPs, many tools and platforms have emerged. As mentioned in **Sec. 7.4**, a notable example is ColabFit Exchange, which functions as an informatics platform tailored for advanced materials and chemistry applications.[134] ColabFit Exchange contains curated, high-quality data from publications useful for fitting MLIPs. As of January 2025, there are nearly 400 datasets comprising more than 230 million unique atomic arrangements. Recent work from Andolina and Saidi generated curated training datasets of 23 single-element systems and built MLIPs with DeepMD, where all of the training data are hosted on ColabFit Exchange.[135,136] Furthermore, packages like the Knowledgebase of Interatomic Models-based Learning-Integrated Fitting Framework (KLIFF) have been developed for general-purpose fitting of MLIPs, offering the versatility to deploy these models within simulation software like LAMMPS via OpenKIM, as well as automated model verification, testing (i.e., the automated computation of a wide range of physical properties for all archived potentials), and archiving on https://openkim.org.[92] KLIFF also incorporates uncertainty quantification, a powerful feature for assessing the reliability and confidence associated with MLIP predictions. These tools exemplify some of the concerted efforts made by the community to surmount adoption barriers and propel the field of MLIPs forward. Another emerging platform in the ecosystem is Garden.[137,138] Garden is designed to make ML models more accessible and deployable across different computing environments. Models are collected into domain-specific "gardens", as a collection of containerized models linked with structured data via the Materials Data Facility[139,140] or Foundry,[141] benchmarks, tests, and computing resources. Garden addresses key infrastructure challenges by containerizing models for consistent execution, facilitating model discovery and simplified deployment across local machines, cloud resources, and HPC clusters through Globus Compute integration. Finally, as discussed in **Sec. 7.4**, at present there are at least two notable examples of interatomic potential repositories, OpenKIM and the NIST Interatomic Potentials Repository, that include many PBPs but also a growing number of MLIPs.

Below, we discuss a standard set of procedures and software for generating or acquiring training data, fitting, comparison, and deployment of MLIPs in MD simulations. For MD simulations, LAMMPS has been a standard in the past decades in the field of materials science with a comprehensive documentation and most widely used MLIPs have libraries in LAMMPS. The earlier MLIP formulations such as Behler-Parrinello NNs, GAP, SNAP, and ACE have well-tested libraries (ML-HDPNN, ML-QUIP, ML-SNAP, and ML-PACE) that have become part of the LAMMPS code and is easier for users to install and use them. MTP also has a LAMMPS library, but it needs to be separately acquired from its Gitlab repository and added to LAMMPS. Recent MLIPs such as DeepMD, MACE and Allegro also have LAMMPS libraries, but currently their libraries need to be downloaded from their GitHub repositories and added to LAMMPS. More streamlined integration of state-of-the-art MLIPs with molecular simulation codes is ongoing. For example, the newest MLIPs (e.g., NequIP, MACE) now provide native integration with JAX-MD, which is a Python library to run end-to-end differentiable MD simulations on GPUs.[142] In addition to LAMMPS and JAX-MD, another Python library frequently used for MD simulations is the Atomic Simulation Environment (ASE).[143] ASE provides numerous functionalities such as MD simulations or static energy/force calculation for each atomic configuration, that can be used for testing and comparing MLIPs. Many aforementioned MLIPs have specific libraries to use with ASE which are called calculators. U-MLIPs such as M3GNet,[41] CHGNet,[20] MACE-MP0,[32] and EquiformerV2-OMAT24 [24] integrate seamlessly with ASE code, enabling a new user to load in a pre-trained U-MLIP and perform atomic relaxations or MD runs with only a few lines of python code.

As discussed in **Sec. 6** and **Sec. 7.1**, many MLIPs require GPUs for efficient operation. Access to modest numbers of GPUs (e.g., 1-10) is becoming widespread in computational labs but can still be challenging to access when many are needed for large-scale studies. The Department of Energy (e.g., Summit) and National Science Foundation (e.g., ACCESS, National Artificial Intelligence Research Resource (NAIRR)) all have machines with large numbers of GPUs to which researchers can apply for resources. In addition, cloud computing resources, e.g., from Google Cloud, Amazon Web Services (AWS) and Microsoft Azure can be leveraged to carry out large simulations with modest cost. This "pay-as-you-go" infrastructure provides users with instant access to state-of-the-art GPUs for large-scale applications. The Garden framework further simplifies access to these diverse computing resources by providing standardized methods for deploying MLIPs across different platforms. Through its integration with Globus Compute, Garden allows researchers to seamlessly utilize various computing resources, from local machines to DOE facilities and cloud providers.

# 9 Limits of Standard MLIPs and Advanced MLIPs to Overcome Those Limits

MLIPs have significantly enhanced our ability to describe PESs in various material systems. When dealing with complexities such as long-range forces, magnetism, and electronic excitation states, it is generally the case that modifications to standard MLIPs are needed. However, adding more physics is more difficult than simply including more data for training MLIPs. In this section, we provide an overview of the limits of MLIP application within the realm of complex materials and the recent advancements to overcome these constraints.

## 9.1 Long-range interactions

Long-range interactions are not included in standard MLIPs as they typically focus on learning local atomic descriptors for environments encompassing a radius of just 5-10 Å, becoming much slower for longer ranges. A graphical depiction of long-range interactions researchers hope to integrate into future MLIPs is given in **Figure 8**. It is possible that the MLIPs which consider contributions only from short-range interactions may be deficient for accurately predicting some properties.[144] In cases where the importance of nonlocal physics and chemistry is fundamental in explaining properties, it becomes imperative to focus on nonlocal electrostatic and dispersion interactions, which are usually not represented by local descriptors. To overcome this challenge, several methodologies and models are employed to enhance the performance of MLIPs for handling long-range interactions.

The first strategy is implicitly incorporating long-range interactions into short-range interactions, which is particularly useful for homogeneous condensed-phase systems with strong screening effects. This essentially comes down to trying to include the correct physics in the training data and hoping the long-range effects are largely screened or reasonably quantitatively renormalized into the short-range MLIP. One approach is increasing the cutoff radius in standard MLIPs to accommodate long-range interactions. For instance, AP-NET utilizes 8 Å cutoff atom-pair symmetry functions for evaluating monomer-monomer interaction energies.[145] A concrete example of renormalizing a naturally long-range interaction is including dispersion in DFT calculations for training data for standard short-range MLIPs. It is interesting to note that for molten salts, which are ionic systems with large electrostatic interactions and significant dispersion contributions, there are many successful MLIP models, demonstrating how effective this simple approach can be.[128,146–148] The ability to represent long-range electrostatics with short-range interactions can be understood as a result of screening, where local charge neutrality makes longer range interactions zero on average. The nature of this screening has been explained and studied quantitatively by Ceder et al.[149] Their work points out that local charge neutrality is strongly correlated with lower-energy states, and that higher-energy states, where local charge neutrality is less robust, have electrostatic interaction that are not well-represented with short-range interactions. Thus, the success of short-range potentials for ionic systems may be in a large part due to the typical states explored in training and application data, which are often lower-energy states associated with near-equilibrium molecular dynamics simulations. These observations imply that for simulations with higher energy states, or more precisely, states without strong local charge neutrality, researchers should be very careful about using only short-range interaction and more complete treatment of long-range electrostatics may be necessary.

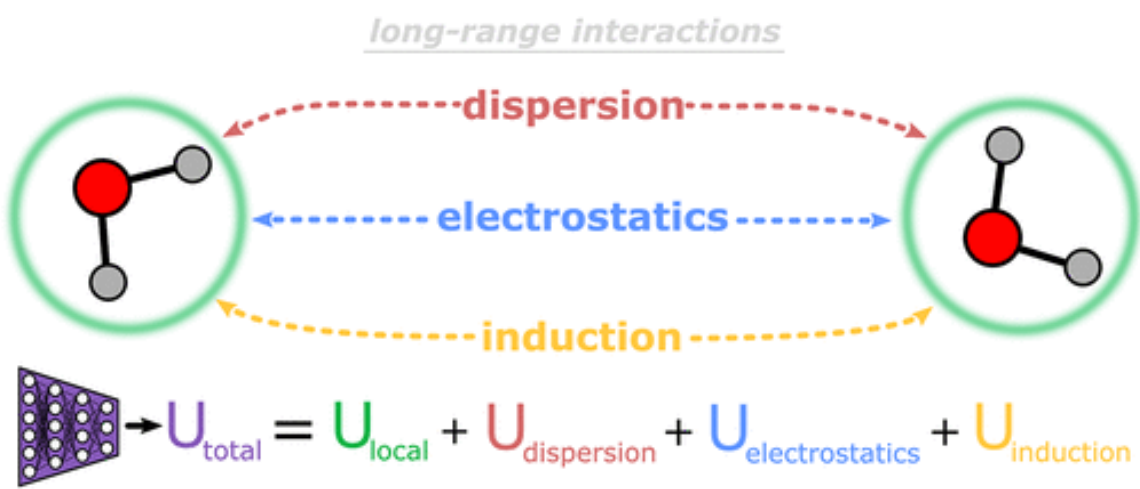

**Figure 8.** Summary of the general energetic contributions composing the total potential energy ($U_{total}$) of a system. $U_{local}$ refers to the short-range system energies and is typically inferred using a machine learning model trained on local features. Dispersion corrections, electrostatics, and induction are collectively referred to as the long-range interaction energy contributions. Adapted with permission from Ref. [144].

The second strategy is including explicit long-range interactions, such as electrostatics, using physics-based functional forms like Coulomb's law, with or without a dependency on the local atomic environment. For instance, the deep neural network potential called DeepPot utilizes a model based on (averages of the positions of) maximally localized Wannier centers to accurately calculate electrostatics.[150] A more refined version of DeepPot is the self-consistent field neural network (SCFNN), where SCFNN combines an iterative refinement approach with maximally localized Wannier centers to enhance the accuracy of electrostatics calculations, demonstrated by its ability to accurately predict the high-frequency dielectric constant of water.[151] The recently updated AIMNet2 (also mentioned in **Sec. 5**) directly includes long-range interactions into the MLIP formalism, in which the DFT-D3 vdW and electrostatic corrections are explicitly included as energy terms, allowing an expanded application to neutral and charged states, as well as diverse organic compounds composed of many different chemical elements.[18,144] Also, as mentioned in **Sec. 5**, the MACE-MP0 U-MLIP was trained only on PBE-level DFT calculations (which only incorporate short-range interactions) and has the ability to add on the DFT-D3 vdW interactions, which are an empirical correction on top of the PBE-level model. This correction can be done easily using the torch-dftd dispersion model implemented in PyTorch.[80] Another example of dispersion corrected-MLIP is SO3LR.[152] As another example, in the global gradient-domain machine learning force field, i.e., Symmetric Gradient Domain Machine Learning (sGDML) approach, the descriptor of a molecular system is treated as a unified entity, bypassing the need for arbitrary partitioning of energy into atomic contributions.[153] The learned model essentially includes all interaction scales and approach enables the sGDML framework to effectively capture both chemical interactions and long-range forces. However, due to the requirement of all correlations of atom-atom interactions, the global MLIPs are trained for a specific molecule and are not transferable to other systems.

In summary, the presence of long-range interactions has posed some challenges for MLIPs, and substantial efforts have been made to address this issue. Specifically, the first strategy of renormalizing long-range interactions from training data into short-range interactions in the MLIP has been extensively employed in standard MLIPs, requiring no additional knowledge or extra effort. For applications where long-range interactions have minimal impacts, users are encouraged to implement this straightforward approach for the first strategy. The second strategy of explicit long-range interactions is becoming more routine for state-of-the-art MLIPs. For studies that require long-range interactions or where such interactions are of interest, particularly electrostatics (e.g., ions, electrolytes), users are encouraged to employ the MLIPs mentioned in the second strategy discussed above. In addition, small molecular systems, usually consisting of (at most) a few hundred atoms, where significant long-range interactions are in play, are well-suited for using global representations of the features. Utilizing a global representation of the entire system typically leads to a reduction in computational complexity compared with previous methods, thereby enhancing both the training process and the efficiency of molecular simulations.

## 9.2 Modeling Systems Off the Born-Oppenheimer Surface

MLIPs are typically a mapping of atomic positions to energy and forces, and therefore assume this mapping is unique. The natural unique PES is that of the lowest energy electron configurations for each atomic arrangement, which is the Born-Oppenheimer surface. However, it is often of interest to consider some forms of excitations. If the excitations are fixed, e.g., we ionize the system, then this is just another uniquely defined Born-Oppenheimer surface determined by some constraint and presents no fundamental challenge. One can simply train a standard MLIP on data from the constrained system Born-Oppenheimer surface. However, if the excitations can move between different Born-Oppenheimer surfaces, e.g., multiple magnetic states or varying electronic excitations, then a significant change in the MLIP formalism is required. Here, we discuss two areas being widely studied, namely magnetism and electronic excitations, although other types of excitations might also be of interest.

### 9.2.1 Magnetism

Different magnetic states of ions possess significantly different properties, and this complexity becomes critical in the context of magnetic materials. How to differentiate ions with different spin states is difficult and lacks a unique solution in the MLIP community. Incorporating spin degrees of freedom into MLIPs, which are crucial for accurately representing finite temperature phenomena in magnetic materials, has remained a challenging task. In spin density functional theory (SDFT), magnetization arises from the interplay between magnetic exchange and band energy contributions,[154,155] where the energy required for electron redistribution between up and down spin channels depends on the local density of states (DOS). Iron, for example, exhibits a bimodal DOS in its body-centered crystal structure (bcc), resulting in larger magnetic moments compared to the face-centered cubic (fcc) structure with a more unimodal DOS.[156] This intricate relationship between magnetic and atomic structure necessitates the consideration of multi-atom, multi-spin interactions to capture various magnetic and atomic arrangements within a single model. Unlike methods derived from electronic structure theory that seamlessly incorporate the complexity of magnetic interactions,[156] classical PBPs require additional terms to mimic quantum exchange interactions. One common approach involves using a classical Heisenberg Hamiltonian,[157] where atomic spin operators are replaced by spin vectors, and exchange interactions are parameterized using *ab initio* calculations.[158] Many MLIP approaches for magnetic systems have adopted similar strategies. For instance, Nikolov et al.[159] expanded the SNAP framework with a two-spin bi-linear Heisenberg model. Yu et al.[160] developed a neural network-based approach to describe contributions to the Heisenberg Hamiltonian based on the local magnetic environment, although this method did not account for lattice information and treated magnetic moments as unit vectors. Eckhoff and Behler[161] extended the original Behler-Parrinello[19] symmetry functions framework but the formalism was limited to collinear configurations. Novikov et al.[162] incorporated magnetic moments as additional degrees of freedom in the MTP framework, albeit also restricted to collinear moments. Domina et al.[163] extended the SNAP framework to handle arbitrary vectorial fields, demonstrating its functionality with non-collinear spin configurations. Chapman and Ma introduced a neural network correction to an embedded atom method potential augmented with a Heisenberg-Landau Hamiltonian for large-scale spin-lattice dynamics simulations.[164] Finally, as discussed in **Sec. 5**, the CHGNet U-MLIP developed by Deng et al.[20] goes beyond reporting energies and forces by also predicting the magnetic moment on every atom in the system, enabling differentiation of different valence states and analysis of the underlying magnetic properties. Despite these efforts, none of the existing ML approaches for magnetic systems have achieved a transferable and quantitatively accurate description of magnetic interactions suitable for modeling magnetism in different crystal structures.

The ACE method has been expanded to accommodate vectorial or tensorial characteristics, alongside the inclusion of atomic magnetic moments and charges in addition to atomic positions, as detailed by Drautz et al.[165] This extended ACE framework offers a complete foundation for characterizing the local atomic environment. Unlike being limited to representing energies solely as a function of atomic positions and chemical species, ACE can be adapted to encompass vectorial or tensorial properties and incorporate additional degrees of freedom. This adaptability is particularly significant for magnetic materials where potential energy surfaces depend on both atomic positions and atomic magnetic moments concurrently. Notably, recent work by Rinaldi et al. introduced a non-collinear magnetic ACE parameterization specifically tailored for the prototypical magnetic element, iron.[166] The model was trained using a diverse set of collinear and noncollinear magnetic structures, computed using SDFT. Their findings demonstrate that this non-collinear magnetic ACE method not only accurately reproduces the ground state properties of various magnetic phases of iron but also captures magnetic and lattice excitations crucial for an accurate description of finite-temperature behavior and crystal defect properties.[166]

Recently, Yu et al.[167] introduced the Time-reversal Equivariant Neural Network (TENN) framework, which incorporates time-reversal symmetry into the equivariant neural network (ENN). This extension allows ENN to account for physical aspects related to time-reversal symmetry, such as the spin and velocity of atoms. Specifically, they developed TENN-e3, an expansion of the E(3) equivariant neural network, to maintain the time reversal E(3) equivariance while considering the inclusion of the spin-orbit effect in situations involving both collinear and non-collinear magnetic moments in magnetic materials. TENN-e3 can construct a spin neural network potential and the Hamiltonian for magnetic materials based on *ab initio* calculations. TENN-e3 employs Time-reversal-E(3)-equivariant convolutions to model interactions between spinor and geometric tensors. TENN-e3 excels at accurately describing the complex spin-lattice coupling while preserving time-reversal symmetry, a feature not present in existing E(3)-equivariant models. Additionally, TENN-e3 facilitates the construction of the Hamiltonian for magnetic materials with time-reversal symmetry.

In summary, TENN offers a new approach for conducting spin-lattice dynamics simulations over extended time scales and performing electronic structure calculations on large-scale magnetic materials. As an instance of TENN-e3, Spin-Allegro can help generate the spin interatomic potential.[168] On the other hand, the ACE approach for iron can be directly extended to multicomponent systems, such as technologically important magnetic alloys and carbides. While conceptually straightforward, generating precise and comprehensive DFT reference data for magnetic multicomponent materials is challenging. People can use efficient sampling techniques based on D-optimality active learning to address this challenge (see **Sec. 7.5.4**), which is expanded to include magnetic degrees of freedom. It can help reduce the number of required DFT reference calculations. Although these novel methods have been proposed, the testing has only been on a small number of systems. Therefore, further exploration and testing of such MLIPs on more magnetic systems are needed to assess the general efficacy.

### 9.2.2 Excited states

At present, a well-established MLIP specifically for excited systems does not exist. Nevertheless, it is crucial to emphasize that ongoing research efforts aimed at developing and enhancing MLIPs are progressing rapidly, and we discuss a few recent efforts here.

Electronically excited states are central to various fields such as photochemistry and photophysics. Like magnetism, they represent an additional degree of freedom that must be added to the potential. Most MLIPs are attempting to learn the PES of molecular/condensed phase systems at the ground state. It requires careful consideration to design an MLIP that can learn the secondary outputs, i.e., excited-state PES, corresponding forces, and nonadiabatic and spin-orbit couplings between them.[169,170] For instance, multiple PESs and their couplings should be considered when dealing with excited states. Furthermore, the complexity and high computational expense of generating the underlying training data calculations and the associated complexity of the corresponding ML models make it more challenging to train an MLIP for excited states than for the ground state. Therefore, the application of ML models for excited states is significantly more challenging than for the ground state.

Recently, Marquet and co-workers developed SchNarc, a framework for excited-state molecular dynamics simulations.[171] SchNarc combines the surface hopping including arbitrary couplings (SHARC) approach for photodynamics, which handles states of different multiplicities, with SchNet (a message-passing deep neural network), which efficiently and accurately fits potential energies and other molecular properties. This framework overcomes current limitations of existing MLIP-based MD simulations for excited states by allowing (i) phase-free training, eliminating the costly preprocessing of raw quantum chemistry data, (ii) treatment of rotationally covariant non-adiabatic couplings (NACs), which can either be trained or (iii) approximated from only ML potentials, their gradients, and Hessians, and (iv) handling of spin-orbit couplings. They extended the model using a NN with multiple outputs to fit all non-adiabatic vectors between different states of the same spin multiplicity simultaneously,[172] which increases the accuracy of the prediction of excited-state dynamics simulations.

More recently, Zhang and co-workers applied a symmetry-adapted high-dimensional neural network to treat couplings as derivatives of NN representations.[173] In this approach, electronic friction was modeled using machine learning and applied to MD simulations of molecules at metal surfaces, thereby treating electron-nuclei coupling in a rotationally covariant manner. For the non-adiabatic coupling vectors, a similar strategy akin to force-only training for potentials, by implementing them as derivatives of virtual properties—properties not explicitly defined in quantum chemistry—constructed by a deep NN. They extended their embedded atom neural network to a universal field-induced recursively embedded atom neural network (FIREANN) by introducing pseudo atomic field vectors relative to each atom with rigorous rotational equivariance. The FIREANN is capable of predicting multiple polarization values for various response properties, making it possible to accurately capture the excited-state PESs within a single model.[174]

## 10 The Future of MLIPs

Given the rapid development and evolution of the field of MLIPs, discussing the future of MLIPs is quite speculative. In particular, the extraordinary pace and disruptive nature of innovations in ML suggest that all predictions related to this area are highly uncertain. With that caveat, we share a few ideas of how the field of MLIPs may progress in the near future.

In the near term (~3-5 years) we see numerous areas where trends that are already well-established are likely to continue. In terms of sampling, we expect to continue to see new methods of active learning and ways to determine optimal training structures to emerge, e.g., as done recently by Fonseca et al. who used ML to sample new areas of configurational space to more optimally improve both explicit AEF MLIPs based on GAP and a deep learning MLIP.[175] Additionally, in complex chemical applications such as bond breaking/formation, advances in active learning can guide the selection of relevant training data, as was illustrated by Kulichenko et al.[124]

At this point there are many different databases available that are used for fitting (e.g., the Materials Project and data shared through ColabFit[134,136]), typically developed by single groups in many different ways. However, there does not seem to be a leading established approach to developing databases of pre-existing *ab initio* calculations for fitting. This problem has many facets as it involves interacting with large existing databases, integrating data from multiple levels of accuracy, and providing guidance for fitting everything from very focused potentials (e.g., just C or Si) to large U-MLIPs (e.g., with 90+ elements). We expect that a few underlying approaches and key databases will eventually become standard and widely adopted for the majority of use cases.

We also expect further refinements to standard MLIP methods. At present, it seems that the pace of innovation has slowed compared to what was occurring over the last 10-15 years, and it appears that the explicit AEF approaches provided by methods like ACE and the deep learning equivariant GNNs are close to optimal within our present understanding. Therefore, within the present explicit AEF and deep learning MLIP framework, efforts will shift to modest changes in the formalism, with a focus on allowing more rapid and turn-key fitting and evaluation of these methods, as well as scaling the fitting to larger datasets. There is a clear need to establish standard methodological approaches to some of the known limitations of present standard MLIPs, which include incorporating long-range-forces and excited states (including magnetic states). Short- and long-range forces are relatively straightforward to treat with either targeted fitting and/or semi-empirical corrections. Recent work has also shown a path for magnetic states, which can be treated by some advanced methods, such as the non-collinear magnetic ACE method and the TENN model, and they are expected to be a standard part of MLIP fitting packages within the next few years. More general excited state methods are being developed and will likely become widely accessible in the next 3-5 years. That all said, the tools of deep learning keep improving, driven by enormous commercial and national priority pressures, which will likely drive rapid improvements in deep learning training, execution, accuracy, interpretation, and implementation. To help readers appreciate the rate of change and improvement in this field, we note that the modern form of the MPGNN upon which so many deep learning MLIPs are based is generally attributed to work published only in 2017.[176] It therefore seems likely that there will be disruptive innovations in ML that will suggest new and possibly much more powerful MLIP approaches sometime within the next 3-5 years, which may alter the focus of the field significantly.

Also in the next 3-5 years, we expect large and crucially important improvements in MLIP-related infrastructure and corresponding increases in the adoption of MLIPs for molecular modeling across the chemistry, materials, physics, and biology communities. Many studies using MLIPs are still related to benchmarking or basic property prediction, and reuse of MLIPs for complex property modeling and materials discovery and design is still limited. However, as their utility becomes better known and the MLIP infrastructure develops further, we can expect much more widespread use. In terms of MLIP infrastructure, code packages for fitting (e.g., MTP, ACE, Allegro, etc.) and integration with major molecular development packages (e.g., ASE, pymatgen) and simulation tools (e.g., LAMMPS) are already widely available, but can still be made more comprehensive and easier to use. Furthermore, greater integration between MLIP fitting codes is likely to provide many advantages. For example, we expect there to soon be code packages that can fit multiple potentials and provide assessment of which is best for your systems and problem. Similarly, such codes and pre-fit MLIPs will be housed in easily accessible and searchable repositories with an automated assessment of MLIP quality, as is being developed in OpenKIM.[177] Both fitting and assessment will greatly benefit from a large set of high-quality benchmark databases. Many benchmarks already exist but were often not developed with MLIP development and benchmarking in mind (e.g., the Materials Project). Applying FAIR principles to MLIPs will increase the useful infrastructure and enhance their adoption. The Garden framework represents an early example of this trend, providing a FAIR-oriented platform that simplifies model publishing, discovery, and deployment across various computing resources. Such frameworks will help democratize access to MLIPs and ensure reproducibility across different computing environments. Finally, we note that direct integration with DFT packages is possible (e.g., as has happened in the VASP code[178] and the Castep code[179]) but that does not seem to be the direction the field is moving, likely due to the ease of connecting DFT with the MLIP fitting and the challenges of maintaining all the advantages of the flexible and evolving MLIP ecosystem when integrated with a DFT package. Overall, navigating the multitude of available options for MLIPs is likely to be a daunting task for at least a few years to come. To facilitate decision-making on the choice of MLIPs for a given application, we expect to see the emergence of recommendation systems based on the user's intended applications and case-specific problems. In this regard, we believe it is important to establish a basis for informed decision-making, i.e., comparisons that aid in evaluating the suitability of different packages.

As an external factor affecting MLIPs, the continually evolving supercomputing landscape can alter the relative strengths of MLIPs based on their ability to adapt. Already, the dominance of GPU-based supercomputers (9 out of the top 10 in the world) renders those MLIPs equipped with GPU-acceleration favorable for scientific applications that require large-scale simulations, as a CPU-locked MLIP will require hundreds of CPU cores to match the performance of even a single GPU. Such differences will be exacerbated as the computing landscape becomes more diversified. Even today, of the four fastest supercomputers, one is CPU-based, while the other three use GPUs from different vendors, whose native programming models are not interoperable. For the typical user with access to one or a small handful of computing environments, the choice of MLIP will strongly be influenced by the MLIP's performance, or even ability to run, on the hardware available to the user. This favors MLIPs that are built on a performance portability layer that makes them largely agnostic to the underlying hardware, such as SNAP and FLARE, which use Kokkos, and many of the deep learning based MLIPs using PyTorch, such as MACE and Allegro. In the future, we may see more radical changes to hardware. Very recently, the Cerebras wafer-scale AI chip was used to run MD simulations more than two orders of magnitudes faster than CPUs and GPUs.[180] While the Cerebras-based simulations used an EAM potential, the results demonstrate the promise of new hardware to drastically change the capabilities of MD simulations, and the MLIPs and their implementations that best adjust accordingly will have a great advantage over the competition.

In the mid-term (5-10 years), a particularly interesting area will be the development of U-MLIPs. U-MLIPs are somewhat analogous to the foundational models that have been so impactful in the image generation and natural language processing (NLP) community. Foundational models generally refer to large models that can achieve good performance on a wide variety of tasks, which allows them to be adapted to many specific applications (i.e., they are a foundation for many other useful more specific models). For example, Large Language Models (LLMs) in the NLP community have seen an explosion of performance and utility over the last few years, and are being integrated into hundreds of different tools and products. The generality of U-MLIPs across chemistry and structure will also allow them to impact many more problems than a typical PBP or targeted MLIP has done in the past, which is why they are sometimes referred to as foundational models for materials and chemistry. At present, U-MLIPs are mostly useful for qualitative or semi-quantitative screening across many systems, but they are rapidly becoming quantitative tools for detailed molecular modeling of specific material properties (e.g., Li diffusion in electrolytes). Future U-MLIPs may function as foundational models, enabling simulation of longer time scales (e.g., > 1 ms) and modeling of totally new materials phenomena currently inaccessible with today's MLIPs. It is likely U-MLIPs will continue to improve rapidly, increasingly taking over the applications presently dominated by targeted MLIPs. This transformation will require a few improvements, but all seem to be well underway. First, larger, more diverse, high-fidelity training data is needed. However, improved hardware, both CPU and GPU, will contribute to increasing the output of *ab initio* data for fitting. Integration of multiple databases will allow for very large training sets and potentially multifidelity training sets[81,181,182] (e.g., with DFT and coupled cluster data) to support MLIPs that approach chemical accuracy (1 kcal/mol) and overcome limitations of lower fidelity DFT data (e.g., like DFT-PBE calculations). We also expect infrastructure and methodological innovations to allow for more contributions from the enormous amounts of data in the broader community, e.g., through online fine tuning or federated learning approaches. It is reasonable to expect that training data sets approaching or exceeding a billion training data points will be within reach in the next few years (we are already seeing training on ~110x10$^6$ DFT configurations). Along with this data, better algorithms and faster GPUs will support more rapid training and evaluation. A final piece that needs to be developed is likely some form of distillation (transferring knowledge from a larger to a smaller model) to allow fast models for specific applications to be easily developed from slower U-MLIPs. This distillation could be as simple as fitting a simpler and faster MLIP to U-MLIP data, but more sophisticated direct methods might be developed. All of these innovations will require improved infrastructure to have their full impact realized. In particular, the scale of data and perhaps even model size of U-MLIPs will require them to be trained and likely hosted by just a few leading organizations with large resources, including perhaps government (e.g., NIST), companies (e.g., Google, Matlantis), major research groups, and relevant societies (e.g., American Chemical Society (ACS)). Such hosting should allow easy use of the models, fine tuning, and distillation for use in high performance applications. These resources could supply full compute environments, just the codes, or some combination. Similar infrastructure is available for LLMs through tools like the OpenAI and HuggingFace APIs, and these tools play an enormous role in supporting the adoption of the LLMs. Frameworks like Garden[137,138] are beginning to lay the groundwork for this future by providing infrastructure that connects models with distributed computing resources and simplifies deployment across different environments – bridging the gap between model developers and users, much like how APIs from OpenAI and HuggingFace have done for LLMs.

As discussed in **Sec. 5**, the change from targeted MLIPs (≤ 5 elements) to U-MLIPs (40-90+ elements) is a continuum. It is possible that semi-universal (SU-MLIPs) (see **Sec. 5**) for key classes of materials with intermediate numbers of elements (e.g., ~20) and/or limited phases or structures, might be established, e.g., for organic molecules, polymers, steel alloys, Al alloys, halide perovskites, electronic materials, molten salts, etc. Such an approach would mimic the very successful methods of the calculation of phase diagrams (CALPHAD) community, which typically develops databases in this manner. Such an approach obviously limits compositional complexity by treating fewer species, and limits the structural complexity by treating fewer phases and structures, but could also make fitting easier by treating relatively consistent physics (e.g., mostly ionic or covalent bonding). Therefore, SU-MLIPs may provide a more practical solution for many materials design problems than full U-MLIPs, or at least bridge the transition from models containing a few species under limited conditions to those seeking to represent the full periodic table under all conditions.

Overall, the above trends will likely lead to a significant reduction in *ab initio* molecular dynamics simulation time, although only after the method has been used to help train many potentials. This reduction may reduce overall compute and energy requirements for molecular modeling research, but we expect a large increase in MLIP modeling, which may offset any gains and likely lead to an increase in the overall utilization of simulation.

More long-term (>10 years), it is possible the traditional potentials (e.g., Lennard-Jones, EAM, AMBER, etc.) will be almost fully replaced by MLIPs, but this is not clear. For example, the AMBER potentials for many organic systems are close to chemical accuracy and very fast, making it unclear what advantages more complex MLIPs would provide. However, it is possible that MLIP approaches will be integrated into even the fastest and simplest potential approaches. For example, Yu et al.[183] recently described an approach to fit pair potentials with ML and then convert them to simple Buckingham form, achieving almost optimal pair potentials from ML with no loss of speed. It is also possible that MLIPs will grow to become much more like full *ab initio* simulations, providing not just a mapping of positions to energies and forces but also to band structures, magnetic moments, charge densities, and even wavefunctions, replacing huge parts of what is presently done with quantum simulations.[184] On the other hand, a complementary vision is that ML integrates with *ab initio* at a more fundamental level, e.g., advancing exchange-correlation functionals and/or massively accelerating solutions of the Schrödinger equation (and relativistic extensions). This path might speed up *ab initio* methods to the level of MLIPs, effectively achieving an MLIP from a very different starting point.[185] Finally, there is perhaps no scientific or engineering field changing as fast as AI and ML right now, so all researchers need to be vigilant for new ideas that can bring entirely new frameworks and capabilities to the molecular modeling community.

# 11 Acknowledgments

Funding for the "Machine Learning Potentials – Status and Future (MLIP-SAFE)" workshop and development of this paper was provided by the National Science Foundation through an AI Institute Planning Grant, Award Number 2020243.

KC thanks the National Institute of Standards and Technology for funding, computational, and data management resources. This work was performed with funding from the CHIPS Metrology Program, part of CHIPS for America, National Institute of Standards and Technology, U.S. Department of Commerce. Certain commercial equipment, instruments, software, or materials are identified in this paper in order to specify the experimental procedure adequately. Such identifications are not intended to imply recommendation or endorsement by NIST, nor it is intended to imply that the materials or equipment identified are necessarily the best available for the purpose.